\documentclass[11pt]{article}
\pdfoutput=1

\usepackage{jheppub} 
\usepackage{amsmath,amssymb,epsfig,amsfonts}
\usepackage{epsf}
\usepackage{makeidx}
\usepackage{slashed}
\usepackage{verbatim} 
\usepackage{float}
\restylefloat{table}
\usepackage{pdflscape}
\usepackage{tikz}
\usepackage{pifont}
\usepackage{array}
\usepackage{youngtab}
\usepackage{multirow}
\usepackage{xcolor}
\usepackage{ulem}
\usepackage{cleveref}
\usepackage{tensor}
\usepackage{todonotes}
\usepackage{mathrsfs}
\usepackage{changepage}
\usepackage{anyfontsize}

\makeatletter

\newcommand{\be}{\begin{equation}}
\newcommand{\ee}{\end{equation}}
\newcommand{\ba}{\begin{aligned}}
\newcommand{\ea}{\end{aligned}}

\newcommand{\gs}{\mathrm{g}_{s}}
\newcommand{\ls}{\ell_{s}}

\newcommand{\dd}{\mathrm{d}}
\newcommand{\me}{\mathrm{e}}
\newcommand{\ii}{\mathrm{i}}
\newcommand{\vol}{\mathrm{vol}}

\newcommand{\lp}{\ell_{p}}
\def\Im{\mathop{\mathrm{Im}}\nolimits}
\def\Re{\mathop{\mathrm{Re}}\nolimits}

\newcommand{\M}{\mathcal{M}}

\renewcommand{\S}{{\rm S}}

\newcount\hour \newcount\minute
\hour=\time \divide \hour by 60
\minute=\time
\def\now{%
\ifnum \hour<13
  \ifnum \hour=0 \advance \hour by 12 \number\hour:\else \number\hour:\fi%
     \ifnum \minute<10 0\fi%
     \number\minute%
\ A.M.%
\else \advance \hour by -12 \number\hour:%
  \ifnum \minute<10 0\fi%
  \number\minute%
  \ P.M.%
\fi%
}

\newcommand{\dD}{\mathrm{D}}
\newcommand{\bc}{\gamma}

\makeatother

\begin{document}

\baselineskip=18pt  
\numberwithin{equation}{section}  
\allowdisplaybreaks  


\thispagestyle{empty}

\vspace*{0cm} 
\begin{center}
{\fontsize{19}{22}\selectfont {{\bf The Near-Horizon Geometry of Supersymmetric\\[10pt]  Rotating AdS$_\textbf{4}$ Black Holes in M-theory}}}
 \vspace*{1cm}
 
\begin{center}

{\fontsize{12.3}{17}\selectfont Christopher Couzens\footnote{c.a.couzens@uu.nl}, Eric Marcus\footnote{e.j.marcus@uu.nl}, Koen Stemerdink\footnote{k.c.stemerdink@uu.nl}, 
 Damian van de Heisteeg\footnote{d.t.e.vandeheisteeg@uu.nl}
}
\end{center}
\vskip .2cm

 \vspace*{0.5cm} 
\emph{ Institute for Theoretical Physics, Utrecht University \\
 Princetonplein 5, 3584 CC Utrecht, The Netherlands}\\
  
 {\tt {}}

\vspace*{0.8cm}
\end{center}

 \renewcommand{\thefootnote}{\arabic{footnote}}

\begin{adjustwidth}{0.26in}{0.26in}

\begin{center} {\bf Abstract } \end{center}

\vspace{0.32cm}

\noindent
We classify the necessary and sufficient conditions to obtain the near-horizon geometry of extremal supersymmetric rotating black holes embedded in 11d supergravity. Such rotating black holes admit an AdS$_2$ near-horizon geometry which is fibered by the transverse spacetime directions. Despite their clear interest to understanding the entropy of rotating black holes, these solutions have evaded all previous supersymmetric classification programs due to the non-trivial fibration structure.
In this paper we allow for the most general fibration over AdS$_2$ with a flux configuration permitting rotating M2-branes. Using G-structure techniques we rewrite the conditions for supersymmetry in terms of differential equations on an eight-dimensional balanced space. The 9d compact internal space is a U$(1)$-fibration over this 8d base. The geometry is constrained by a master equation reminiscent of the one found in the non-rotating case. We give a Lagrangian from which the equations of motion may be derived, and show how the asymptotically AdS$_4$ electrically charged Kerr-Newman black hole in 4d $\mathcal{N}=2$ supergravity is embedded in the classification. In addition, we present the conditions for the near-horizon geometry of rotating black strings in Type IIB by using dualities with the 11d setup.

\end{adjustwidth}
\noindent

\newpage


\tableofcontents
\printindex
\setcounter{footnote}{0}
\newpage 

\section{Introduction}

The idea of extremization principles playing a fundamental role in physics has a long history since the advent of the Lagrangian and the principle of least action. More recently extremal problems have also been shown to play a role in both quantum field theory and supergravity. On the field theory side $a$-maximization \cite{Intriligator:2003jj}, $F$-maximization \cite{Jafferis:2010un}, $c$-extremization \cite{Benini:2012cz,Benini:2013cda} and $\mathcal{I}$-extremization \cite{Benini:2015eyy} have been successfully used to compute observables in SCFTs in 4, 3, 2 and 1 dimension(s) respectively. Via AdS/CFT it is natural to conjecture that there are dual extremization principles on the gravity side. Indeed such geometric extremization principles have been found for all of the field theory principles mentioned above. In \cite{Martelli:2005tp,Martelli:2006yb} a geometric dual to $a$-maximization and $F$-maximization was given whilst in \cite{Couzens:2018wnk} an analogous proposal for $c$-extremization and $\mathcal{I}$-extremization was given for certain classes of theories. The classes of solutions tackled in \cite{Couzens:2018wnk} and in the later works \cite{Gauntlett:2018dpc,Hosseini:2019use,Gauntlett:2019roi,Hosseini:2019ddy,Kim:2019umc, Gauntlett:2019pqg,vanBeest:2020vlv} are AdS$_3$ solutions in Type IIB and AdS$_2$ solutions in 11d supergravity. Subclasses of these arise as the near-horizon of static black strings and black holes embedded in the respective theories.\footnote{See also \cite{Lozano:2020txg} for an extremization principle for $\mathcal{N}=(0,4)$ AdS$_2$ solutions in Type IIB.}

For example, the near-horizon limit of a static asymptotically AdS$_4$ extremal black hole in 4d gauged supergravity contains an AdS$_2$ factor, see the review \cite{Kunduri:2013gce} and references therein. The staticity of the black hole requires that the transverse directions of the geometry are not fibered over AdS$_2$ but merely form a warped product. If one further restricts to magnetically charged black holes and uplifts the near-horizon solution to 11d supergravity, one obtains a supersymmetric solution with an AdS$_2$ factor and electric four-form charge. Solutions of this form were classified in \cite{Kim:2006qu} and later extended in \cite{Donos:2008ug,Hong:2019wyi} to include additional magnetic flux. The geometries are a warped product of AdS$_2$ with a nine-dimensional internal manifold which is locally a U$(1)$ bundle over a conformally K\"ahler space. To construct these geometries one places M2-branes in an asymptotic geometry of $\mathbb{R}\times CY_5$ and wraps them on a curve inside the Calabi--Yau five-fold. The near-horizon of this setup then gives rise to the AdS$_2$ geometry which in turn is seen to be the near-horizon of a black hole.

In order to obtain an AdS$_2$ solution it was important that the 4d black hole was both static and only magnetically charged. Adding rotation to the four-dimensional black hole leads to the internal space being fibered over the AdS$_2$ in the near-horizon, which will clearly persist in the uplift. Though not as obvious, if the 4d black hole has electric charges which are identified as arising from gauged flavour symmetries, this will also lead to a fibered AdS$_2$ in the 11d uplift. A gauge field in the truncation can have two sources, either it comes from gauging an isometry of the compactification manifold, or from the expansion of a $p$-form potential on $(p-1)$-cycles of the compactification manifold. The former gauge fields are dual to flavour symmetries whilst the latter are dual to baryonic symmetries. For the flavour symmetries the uplift will lead to the isometries being fibered over AdS$_2$ in the 11d solution. In summary, in order to incorporate more general black holes which rotate and have electric charges, one \emph{must} relax the product structure of the 11d solution and allow for the internal manifold to be fibered over AdS$_2$. In contrast, one of the essential ingredients used in the works \cite{Kim:2006qu,Hong:2019wyi}, and more generally in AdS classifications, is that the AdS factor is a direct product in the metric.

In this paper we will lay the groundwork for extending the geometric dual of $\mathcal{I}$-extremization and $c$-extremization to theories arising from the near-horizon of rotating black holes and black strings respectively. Concretely we will classify all supersymmetric solutions of 11d supergravity containing an internal manifold arbitrarily fibered over AdS$_2$. With such a general ansatz we cover the black holes considered in \cite{Nian:2019pxj,Choi:2018fdc,Hosseini:2019iad,Hosseini:2019lkt,Bobev:2019zmz}. To the best of our knowledge this is the first time in the literature this has been performed.\footnote{There is the nice paper \cite{Katmadas:2015ima} where the embedding of static AdS$_4$ black holes in 11d supergravity were considered.} We find that the 9d internal manifold is a U$(1)$ fibration over an 8d space admitting a balanced metric. The balanced metric satisfies a master equation which is the analogue of the one found in the non-rotating case \cite{Kim:2005ez, Kim:2006qu}, see also \cite{Donos:2008ug,Couzens:2017nnr,Passias:2019rga,Couzens:2019iog} for further generalizations of these master equations. Through dualities we also classify a class of rotating black string near-horizons in Type IIB.

The outline of this paper is as follows. In section \ref{sec:setup} we study the necessary and sufficient conditions for a supersymmetric solution with time fibered over the transverse directions and consistent with preserving an SO$(2,1)$ symmetry. In section \ref{sec:action} we give an action from which the equations of motion found in section \ref{sec:setup} may be derived. In particular we show that when supersymmetry is imposed on the action it reduces to a simple form which computes the entropy of the black hole/string. By way of exposition we show in section \ref{Sec:Example} how the electrically charged AdS$_4$ Kerr-Newman black hole is embedded in the classification. Section \ref{sec:BlackString} discusses the conditions on the geometry of rotating black strings in Type IIB by using dualities with the 11d geometry. We conclude in section \ref{sec:conclusion}. A discussion on general black hole near-horizons and computing observables of the solutions is presented in appendix \ref{app:NH}.


\section{Setup}\label{sec:setup}

In this section we will explain the general procedure for obtaining the conditions for preserving supersymmetry of near-horizon solutions of rotating black holes. In general the conditions we find are necessary and sufficient conditions that must be satisfied by the near-horizon of any rotating black hole in 11d supergravity arising from rotating M2-branes. We will determine these conditions by using the results in \cite{Gauntlett:2002fz} which classified all 11d supergravity backgrounds preserving supersymmetry and admitting a timelike Killing vector. Using \cite{Gauntlett:2002fz} we can reduce the 11d supersymmetry conditions into differential conditions on a 10d base space. This base space must be non-compact and upon imposing the natural condition that the 10d space is a cone we can reduce the conditions further to a compact 9d base, $Y_9$. This 9d base is a U$(1)$ fibration over an 8d base, $\mathcal{B}$. In general the 8d base is not conformally K\"ahler, which is true for the non-rotating AdS$_2$ case studied in \cite{Kim:2006qu}, but instead is a conformally balanced space. 

One of the guiding principles that we will use is to impose that the near-horizon solution possesses an SO$(2,1)$ symmetry dual to the conformal group in the 1d superconformal quantum mechanical theory. Generally the ansatz that we will use when reducing the supersymmetry conditions does not possess this full symmetry but only a subset of it. However, from the point of view of imposing supersymmetry it is more convenient to work with this more general setup and then further constrain the geometry to preserve the full conformal group later. We will find that the additional constraints that we need to impose for the existence of an SO$(2,1)$ symmetry are specified by giving a constant vector with entries corresponding to each of the Killing vectors of the metric. These constants are related to the near-horizon angular velocities of the black hole along the Killing directions. 

We begin this section by reviewing the conditions for a supersymmetric geometry in 11d supergravity to admit a timelike Killing vector following \cite{Gauntlett:2002fz}. We discuss in detail the ansatz we will use in performing the reduction and subsequently reduce the conditions to an 8d base space. Up until this point we have not imposed the existence of an SO$(2,1)$ symmetry and in the final part of this section we discuss the additional constraints one must impose for such a symmetry using the results in appendix \ref{app:NH}.

\subsection{Timelike structures in 11d supergravity}
\label{sec:summaryGP}

In \cite{Gauntlett:2002fz} the conditions for a solution of 11d supergravity to admit a timelike Killing spinor were derived. Here we summarize the most important results for our purposes. The metric takes the general form
\be
\dd s^{2}_{11}= -\Delta^{2} (\dd t+ a)^2 + \Delta^{-1} \me^{2\phi} \dd s^{2}_{10}
\ee
where $\Delta$ and $\me^{2\phi}$ are functions defined on the 10d base. Note that we use a rescaling $\me^{2\phi}$ of the 10d metric compared to \cite{Gauntlett:2002fz}. The 10d base admits a canonical SU$(5)$ structure which we denote by $(j,\omega)$\footnote{In comparison to \cite{Gauntlett:2002fz} one should identify $(a,\me^{2\phi}j,\me^{5\phi}\omega,\me^{5\phi}\Re[\omega])_{\text{here}} \leftrightarrow (\omega, \Omega, \theta, \chi)_{\text{there}}$. In particular this transforms the torsion modules as $(\me^{\phi}w_{1},\me^{3\phi}w_{2},\me^{2\phi}w_{3},w_{4}+8\,\dd \phi,w_{5}-40\,\dd \phi)_{\text{here}} \leftrightarrow (w_{1},w_{2},w_{3},w_{4},w_{5})_{\text{there}}$.}. We normalize this structure such that
\begin{equation}
\omega\wedge\bar{\omega} = (-2i)^5\, \frac{j^5}{5!} \,.
\end{equation}
The exterior derivatives of the structure forms satisfy
\begin{align}
\dd j &= \frac{1}{8} w_1 \lrcorner \Im[\omega] + w_3 + \frac{1}{4} w_4 \wedge j \,, \label{JSU5}\\
\dd \Re[\omega] &= \frac{1}{3} w_1 \wedge \frac{j^{2}}{2!} + w_2 \wedge j -\frac{1}{8} w_5\wedge\Re[\omega] \,. \label{OmegaSU5}
\end{align}
Here the $w_{i}$ are the torsion modules of the SU$(5)$ structure: $w_1$ is a real $(2,0)$+$(0,2)$-form, $w_{2}$ a real primitive $(3,1)$+$(1,3)$-form, $w_3$ a real primitive $(2,1)$+$(1,2)$-form and $w_4$ and $w_5$ are real one-forms. The 11d four-form flux is decomposed into 10d fluxes as
\begin{equation}
\mathcal{G}_4 = (\dd t+a) \wedge f_{3}+ h_4\,.
\end{equation}

Following the results of \cite{Gauntlett:2002fz}, imposing supersymmetry yields the following conditions relating the fluxes to the structure forms
\begin{align}
\dd(\me^{2\phi}j) &= f_3 \,, \\
\dd \big(\Delta^{-3/2} \me^{5\phi}\Re[\omega]\big) &= \me^{2\phi}\star_{10}h_4- \me^{2\phi}\, h_4 \wedge j - \me^{4\phi}\, \dd a \wedge \frac{j^2}{2} \,.
\end{align}
Moreover it follows that the 11d flux takes the form
\begin{equation}\label{11dflux}
\begin{aligned}
\mathcal{G}_4 =\: &(\dd t + a) \wedge \dd(\me^{2\phi} j) - \Big[ \frac{3}{4}\, \dd a^{(0)} j + \dd a^{(2,0)}+\dd a^{(0,2)} + \frac{1}{3} \,\dd a^{(1,1)}_{0}\Big]\wedge \me^{2\phi}j \\
& +\frac{1}{2}\me^{-2\phi}\star_{10} \dd \big( \Delta^{-3/2}\me^{5\phi} \Re[\omega]\big)-\frac{1}{2} \me^{-2\phi}\star_{10}\Big[ j \wedge \dd \big(\Delta^{-3/2} \me^{5\phi}\Re[\omega] \big)\Big] \wedge j\\
& -\frac{1}{16}\Delta^{-3/2} \me^{3\phi}\star_{10} ( [w_{5}+4 w_{4}-8\dd\phi] \wedge \Re[ \omega])+ {h}^{(2,2)}_{0} \,,
\end{aligned}
\end{equation}
where $\dd a$ decomposes as $\dd a=\dd a^{(0)} j +\dd a_0^{(1,1)}+\dd a^{(2,0)}+\dd a^{(0,2)}$, and ${h}_{0}^{(2,2)}$ is the primitive $(2,2)$ part of $h_4$ and is unconstrained by supersymmetry. Additionally the torsion module $w_5$ is fixed by supersymmetry to be
\begin{equation}\label{torsionw5}
w_5 = -12\,\dd\log\Delta + 40\,\dd\phi \,.
\end{equation}
For a supersymmetric solution to exist these conditions must be supplemented by the Bianchi identity and Maxwell equation
\begin{align}
\dd h_4 &= -\dd a \wedge \dd(\me^{2\phi}j) \,, \label{bianchi1}\\
\dd \big(\Delta^{-3}\me^{4\phi} \star_{10} \dd (\me^{2\phi}j)\big) &= \me^{2\phi}\,\dd a \wedge \star_{10}\, h_4 + \frac{1}{2} h_4 \wedge h_4 \,. \label{maxwell11}
\end{align}
The set of equations as given above are both necessary and sufficient for a solution to admit a timelike Killing spinor.

Our main motivation is to obtain the near-horizon geometries of rotating M2-branes wrapped on Riemann surfaces, which may give rise to the near-horizon of rotating black holes. We must therefore make some assumptions about the form of the solution. To engineer such solutions one should place the rotating M2-branes in an asymptotic geometry of the form $\mathbb{R}_t\ltimes CY_5$ and then wrap the M2-brane on a Riemann surface inside the Calabi--Yau five-fold. Note that the rotation of the M2-brane leads to the non-trivial fibration of the 11d space-time, with the time direction fibered over the five-fold. Since the asymptotic geometry is Calabi--Yau it is natural to expect that our 10d base space is complex, which requires that $w_1 = w_2 = 0$. This is indeed how the rotating M2-brane solution is embedded in the classification of \cite{Gauntlett:2002fz}. We will be satisfied with using the complex condition as a well-motivated ansatz in the following though it would certainly be interesting to lift this restriction. In addition to requiring the complex condition we also want to eliminate the possibility of having flux sourcing M5-branes. For this reason we will remove any terms appearing in the flux which are of Hodge type $(4,0)$+$(0,4)$, since these would not come from to M2-branes wrapped on a Riemann surface.\footnote{Lifting this assumption would open up the possibility of studying the near-horizon of rotating asymptotically AdS$_7$ black holes \cite{Cvetic:2005zi} arising from wrapping M5-branes on SLAG five-cycles in the Calabi--Yau five-fold. This would give the 11d geometric setting for the computations performed in \cite{Hosseini:2018dob,Kantor:2019lfo,Benini:2019dyp,Bobev:2019zmz}.} From \eqref{11dflux} and \eqref{torsionw5} we see that this assumption implies $w_4 = 3\,\dd\log\Delta - 8\,\dd\phi$.

Under these assumptions the 10d torsion conditions are
\begin{equation}\label{eq:10dtorsion}
\begin{aligned}
\dd(\me^{2\phi}\, j) &= f_3 \, , \\
\dd(\Delta^{-3}\me^{8\phi}\, j^4) &= 0 \, , \\
\dd(\Delta^{-3/2}\me^{5\phi}\, \omega) &= 0 \, . \\
\end{aligned} 
\end{equation}
The last unspecified torsion module is given by the primitive part of the three-form flux: $w_3 = \me^{-2\phi}\,f_{3,0}$. The 11d flux can now be succinctly written as
\begin{align}
\mathcal{G}_4 &= (\dd t + a) \wedge \dd( \me^{2\phi}j) - \Big[ \frac{3}{4}\, \dd a^{(0)} j + \dd a^{(2,0)}+\dd a^{(0,2)} + \frac{1}{3}\, \dd a^{(1,1)}_{0}\Big]\wedge \me^{2\phi} j + {h}_{0}^{(2,2)}\, \nonumber \\
&= -\,\dd \big[(\dd t+a)\wedge \me^{2\phi} j \big]+\tilde{h}^{(2,2)}\, , \label{eq:fluxreduced}
\end{align}
where we define the shifted four-form flux
\begin{align}
\tilde{h}^{(2,2)} &= h_4 + \dd a \wedge \me^{2\phi}j \,, \nonumber\\
&= {h}^{(2,2)}_0 + \frac{1}{2}\,\me^{2\phi}\,\dd a^{(0)}\, \frac{j^2}{2!} + \frac{2}{3}\,\me^{2\phi}\,\dd a^{(1,1)}_0 \wedge j \,.\label{h22def}
\end{align}
The Bianchi identity \eqref{bianchi1} and Maxwell equation \eqref{maxwell11} can now be rewritten in terms of $\tilde{h}^{(2,2)}$ as
\begin{align}
\dd \tilde{h}^{(2,2)} &= 0 \,, \label{eq:10dbianchi}\\[4pt]
\dd\big(\Delta^{-3} \me^{4\phi} \star_{10} \dd (\me^{2\phi}j) \big) &= \frac{1}{2} \tilde{h}^{(2,2)} \wedge \tilde{h}^{(2,2)} \,. \label{eq:10dmaxwell}
\end{align}
For future reference, we give a few useful identities containing $\tilde{h}^{(2,2)}$:
\begin{align}
\star_{10}\, \tilde{h}^{(2,2)} &= \tilde{h}^{(2,2)} \wedge j - \me^{2\phi}\,\dd a^{(0)} \, \frac{j^3}{3!} -2\,\me^{2\phi}\,\dd a_0^{(1,1)} \wedge \frac{j^2}{2!} \,,\\[4pt]
j \lrcorner\, \tilde{h}^{(2,2)} &= 2 e^{2\phi}\,\dd a^{(1,1)} \label{eq:traceflux}\,.
\end{align}
Here $\dd a^{(1,1)} = \dd a^{(0)} j +\dd a_0^{(1,1)}$, i.e. we omit the $(0,2)$ and $(2,0)$ contributions.

\subsection{Ansatz}\label{sec:ansatz}

To proceed we must now insert an ansatz for the 10d base space. It was shown in \cite{Gauntlett:2002fz} that the base is necessarily non-compact (the argument uses some smoothness conditions but these should hold in the present setting), and so we impose that the base is conformally a cone. The metric we take is
\be\label{ansatzmetric}
\dd s^2_{11} = -\Delta^2 (\dd t +a)^2 + \Delta^{-1} \me^{2 \phi}\big( \dd r^2 +r^2 \dd s_{9}^2\big)\, .
\ee
Next we need to specify how the scalar fields $\Delta,\phi$, connection one-form $a$ and fluxes scale with respect to the radial coordinate. Ultimately we want to be able to recover a warped AdS$_2$ factor and an $r$-independent 9d space. This fixes the scaling of $\Delta$ and $\phi$ to be
\be\label{DeltaAndPhiScaling}
\Delta=\frac{\me^{B+C}}{r}\, ,\qquad\quad \me^{2 \phi} = \frac{\me^{3 B+C}}{r^3}\, ,
\ee
where we have introduced two new scalars $B$ and $C$ which are independent of the radial coordinate. For general scalar $C$ this will not lead to a geometry admitting an SO$(2,1)$ isometry generating the conformal group in 1d. As discussed earlier one must impose additional constraints. Rather than imposing them now it is more convenient to impose them later and leave the scalar $C$ unconstrained for the moment. 

The conical geometry naturally gives rise to an R-symmetry vector $\xi$ defined by
\be
\xi = j \cdot (r \partial_{r})  \,.
\ee 
As can be easily checked by explicit computation the norm squared of the vector is $r^2$. On the link of the cone at $r=1$ this translates to the existence of a unit-norm vector generating a holomorphic foliation over an 8d base admitting an SU$(4)$ structure inherited from the parent SU$(5)$ structure. We denote this 8d base by $\mathcal{B}$. Introducing coordinates for this vector
\be
\xi= \partial_{z}\, ,
\ee
we can write the dual one-form as
\be
\eta = \dd z + P\, ,
\ee
where $P$ is a one-form on $\mathcal{B}$. We may now decompose the SU$(5)$ structure $(j,\omega)$ in terms of the SU$(4)$ structure, which we denote by $(J,\Omega)$, as
\be
\begin{aligned}\label{structureansatz}
j &= r \eta \wedge \dd r + r^{2}\,\me^{-3B-C/3}\, J \, ,\\[2pt]
\omega &= r^4 \,\me^{-6B-2C/3}\, \me^{\ii z} (\dd r- \ii r \eta) \wedge \Omega \, .
\end{aligned}
\ee
Here we include a scaling $\me^{-3B-C/3}$ of the 8d base, and a phase along the $z$-direction. The choice of scaling has been chosen so that the two form is balanced rather than conformally balanced as will become clear in the following section. While the phase is required by supersymmetry and implies that the holomorphic volume form has unit charge under the vector $\xi$. 

The scaling of the connection one-form appearing in the time-fibration is fixed to be
\be\label{ansatzfibration}
a = r(\alpha \,\eta + A)\, , \\
\ee
where $\alpha$ and $A$ denote an 8d scalar and one-form respectively. Note that we did not include a term with a leg on $\dd r$ in this decomposition because such a term could be absorbed by redefinitions and coordinate changes for a near-horizon geometry. It will turn out that imposing the SO$(2,1)$ symmetry will further constrain the one-form $a$ and scalar $C$ however we postpone this discussion to later. The field strength $\dd a$ is
\begin{equation}\label{eq:da}
\dd a = \alpha\, \dd r \wedge \eta + \dd r \wedge A - r\,\eta\wedge \dd\alpha + r\,(\alpha\,\dd\eta + \dd A) \,.
\end{equation}
With these ans\"atze the 11d metric becomes
\be
\dd s^2_{11}= \me^{2 B} \Big[ -\me^{2 C} \Big( \frac{\dd t}{r}+ \alpha\, \eta + A\Big)^2+ \frac{\dd r^2}{r^2}+ \eta^2 + \me^{-3B-C/3}\, \dd s^2_{8} \Big]\, .
\ee
We recover the non-rotating case by setting $\alpha=0$, $A=0$ and $\me^{2C}=1$.\footnote{In comparison to \cite{Kim:2006qu} we identify $B_\text{here} \leftrightarrow A_\text{there}$, and comparing with \cite{Gauntlett:2007ts} we identify $B_\text{here} \leftrightarrow -B_\text{there}/3$.}

Finally we must fix the $r$-scaling of the flux. The scaling is fixed by regularity as $r\rightarrow 0$ and preserving the SO$(2,1)$ symmetry which requires the radial dependence to only appear in the one-forms
\be
\frac{\dd t}{r} \quad \text{and} \quad \frac{\dd r}{r}\, .
\ee
It follows that the 10d fluxes $f_{3}$ and $\tilde{h}^{(2,2)}$ decompose in terms of 8d fluxes as
\begin{align}\label{Hfluxin8Dcomponents}
f_{3} &= r^{-1}\Big[ \frac{\dd r}{r} \wedge \eta \wedge F_{1} + \frac{\dd r}{r} \wedge F_{2} + F_{3}\Big]\, , \\[4pt]
\tilde{h}^{(2,2)} &= H^{(2,2)}+ \frac{\dd r}{r} \wedge ( H^{(2,1)}+H^{(1,2)}) + \eta \wedge \ii (H^{(2,1)}-H^{(1,2)})+\frac{\dd r}{r} \wedge \eta \wedge H^{(1,1)}\, .
\end{align}
In principle one could include a piece of $f_3$ with one leg on $\eta$ and two legs on $\mathcal{B}$, but we omit it here because it will be put to zero by supersymmetry. Note that we keep track of the Hodge type of the components of $\tilde{h}^{(2,2)}$, where the holomorphic and anti-holomorphic one-form associated with $\dd r$ and $\eta$ are given by $e^{1} = \dd r-\ii r\eta$ and its conjugate respectively.

\subsection{8d supersymmetry conditions}\label{sec:8d susy}

We can now derive the 8d conditions by reducing their 10d counterparts using the ans\"atze presented in the previous section. Let us begin by reducing the SU$(5)$ structure torsion conditions to SU$(4)$ structure conditions. From decomposing \eqref{eq:10dtorsion} we find
\begin{align}
F_{1} &= -\,\dd \me^{3B+C}\, , \label{eq:fluxF1}\\
F_{2} &= \me^{3B+C} \,\dd \eta - \me^{2C/3}\,J \, , \\ 
F_{3} &= \dd(\me^{2C/3}\,J)\, , \\
\dd {J^{3}}&=0 \, , \label{eq:balance}\\
\dd \eta \wedge \frac{J^{3}}{3!} &= \me^{-3B-C/3}\, \frac{J^{4}}{4!}\, , \label{eq:tracedeta}\\[3pt]
\dd \me^{-3B-C/3} \wedge {J^{4}} &= 0\, , \label{eq:scalarsconstr}\\
\dd \Omega &= \ii \big( P + \tfrac{1}{3}\dd^{c} C \big) \wedge \Omega \, . \label{eq:ricciform}
\end{align}
Recall that $\Omega$ has unit charge under the vector $\partial_{z}$ which is evident from \eqref{structureansatz}. From these equations we can deduce the SU$(4)$ torsion modules $W_{i}$. From \eqref{eq:ricciform} we immediately see that the 8d base is complex: $W_{1}=W_{2}=0$. Furthermore, from \eqref{eq:balance} we see that $W_4=0$, i.e. the base is balanced. Fixing the two-form to be balanced as opposed to conformally balanced fixed the choice of scaling of the 8d base in \eqref{structureansatz}. In particular the base is not K\"ahler: the third torsion module is related to the primitive part of $F_3$ as $W_3 = \me^{-2C/3}\, F_{3,0}$. However, for the Kerr--Newman electrically charged black hole that we consider in section \ref{Sec:Example} this part of the flux vanishes, and the 8d base is therefore K\"ahler. From \eqref{eq:ricciform} we find $W_5 = -4J\cdot P - \tfrac{4}{3}\dd C$ and this fixes the Ricci-form of the base in terms of the connection $P$ and the scalar $C$ as we show below. Before proceeding it is useful to rewrite the three-form flux $f_3$ as
\be
f_{3}= r^{-1} \Big[ \frac{\dd r}{r} \wedge \hat{F}- \dd \hat{F}\Big]\, , \quad \text{where} \quad  \hat{F}=- \me^{2C/3} J + \dd (\me^{3B+C} \eta)
\ee
which puts it into a form more reminiscent of the non-rotating case \cite{Kim:2006qu}.

Let us turn our attention to the other identities following from \eqref{eq:fluxF1}-\eqref{eq:ricciform}. Firstly, from \eqref{eq:scalarsconstr} we find
\begin{equation}
\mathcal{L}_{\xi} \me^{-3B-C/3} = 0 \, .
\end{equation}
In fact, we will take $\xi$ to be a symmetry of each of the scalars $B,C$ individually, though supersymmetry does not require this. This assumption is natural since we want $\xi$ to play the role of the R-symmetry vector of the solution. Note that these conditions imply that it is a Killing vector of the 10d space and by imposing $\mathcal{L}_{\xi}\alpha=0$, it is in fact a Killing vector for the full 11d metric. Taking the exterior derivative of \eqref{eq:ricciform} implies
\begin{equation}
\dd \eta \wedge \Omega = 0 \label{eq:11form}\, ,
\end{equation}
hence $\dd\eta$ is a $(1,1)$-form on the base. Moreover from \eqref{eq:tracedeta} we find that
\begin{equation}\label{Jintodeta}
J \lrcorner\, \dd \eta  = \me^{-3B-C/3}\, .
\end{equation}
Finally from \eqref{eq:ricciform} we can read off the Ricci form $\rho$ on the 8d space to be
\begin{equation}\label{ricciform}
\rho = \dd \eta + \tfrac{1}{3}\dd \dd^{c} C\,.
\end{equation}
Note that the second term is exact since we require the scalar $C$ to be globally well-defined.

This in turn allows us to compute the Chern--Ricci scalar\footnote{Here we find the d'Alembertian operator through the short computation: $$J\lrcorner\,\dd \dd^{c} C = \ast (\dd \dd^{c} C \wedge \ast J) = \ast\, \dd \,(\dd^{c} C \wedge \ast J) = \ast\,\dd\ast(\dd^c C \lrcorner\,J) = - \ast\dd\ast\dd C = -\Box C\,,$$where we use that $\tfrac{1}{3!}\dd J^3 = \dd\ast J = 0$.}
\begin{equation}\label{chern-ricci scalar}
R_{C} \equiv 2 J \lrcorner\, \rho = 2e^{-3B-C/3}- \tfrac{2}{3} \Box C \,.
\end{equation}
The Chern--Ricci scalar is related to the more common 8d Ricci scalar via\footnote{Note that this is equivalent to the identity $R_8 = R_{C}-\tfrac{1}{2} | \dd^c J |^{2}$ that is also used in the literature.}
\begin{equation}
R_8 = R_{C}-\tfrac{1}{2} | \dd J |^{2}\, .
\end{equation}
It is clear from the above relation that the two scalars coincide when the manifold is K\"ahler. 

So far we have only imposed supersymmetry and not the equations of motion. Integrability of the Killing spinor equations implies that the Einstein equations are satisfied so long as the Bianchi identity \eqref{eq:10dbianchi} and Maxwell equation \eqref{eq:10dmaxwell} are imposed. Imposing these gives us additional constraints on the geometry and fluxes. From reducing the Bianchi identity we find
\begin{equation}\label{8dbianchis}
\begin{aligned}
\dd H^{(2,2)} &= -\ii \dd \eta \wedge (H^{(2,1)}-H^{(1,2)})\, , \\
\partial H^{(2,1)} &= \bar{\partial} H^{(1,2)} = 0 \, ,\\
\bar{\partial}H^{(2,1)} &= \partial H^{(1,2)} = -\tfrac{1}{2}\,\dd\eta\wedge H^{(1,1)} \, , \\
\dd H^{(1,1)} &= 0\, .
\end{aligned}
\end{equation}
From this decomposition it is simple to show that the  R-symmetry vector $\xi$ is not just a symmetry of the metric, but also for the 10d flux $\tilde{h}^{(2,2)}$, i.e.
\begin{equation}
\mathcal{L}_\xi \tilde{h}^{(2,2)}=0\, .
\end{equation}
In fact, we find that $\xi$ is a symmetry for the full 11d flux $G_4$ as well, since by using \eqref{eq:da} and that the scalar $\alpha$ has vanishing Lie-derivative along $\xi$ one can show that
\begin{equation} 
\mathcal{L}_{\xi} G_4=0 \, .
\end{equation} 
This is then consistent with our interpretation of $\xi$ as being the Killing vector dual to the R-symmetry of a putative dual field theory. 

From the 10d Maxwell equation we find the set of equations
\begin{align}
- \dd \ast_{8} \dd  \me^{-3B-C} + \me^{-2C/3}\, \dd\eta \wedge \dd\eta \wedge \frac{J^2}{2!} &=  \frac{1}{2}H^{(2,2)}\wedge H^{(2,2)} \,,\label{maxwell1}\\[2pt]
\me^{-4C/3}\, \dd \eta \wedge  \ast_{8}\, \dd (\me^{2C/3}J) &= H^{(2,2)} \wedge (H^{(2,1)}+H^{(1,2)})\, , \\
- \dd \eta \wedge \dd \Big( \me^{-2C/3}\, \frac{J^{2}}{2!} \Big) &= H^{(2,2)} \wedge \ii(H^{(2,1)}-H^{(1,2)}) \, ,\\[2pt]
- \dd \dd^c \Big(\me^{-2C/3} \frac{J^2}{2!}\Big) &= H^{(2,2)}\wedge H^{(1,1)} + 2 \ii H^{(2,1)}\wedge H^{(1,2)}\, .\label{maxwell4}
\end{align}
It can be shown that the second and third equation are equivalent by acting with the operator $J\cdot$ which acts by contracting the complex structure into each index of the form. For a $(p,q)$-form this acts by multiplying the form by $\ii^{p-q}$. By applying the 8d Hodge star to \eqref{maxwell1}, and by inserting \eqref{ricciform} and \eqref{chern-ricci scalar}, we can rewrite it as
\begin{equation}\label{eq:box}
-\me^{2C/3}\Box\big(\me^{-2C/3}(R_C + \tfrac{2}{3}\Box C)\big) + \tfrac{1}{2}\big(R_C + \tfrac{2}{3}\Box C\big)^2 - 2\big|\rho - \tfrac{1}{3}\dd\dd^c C\big|^2 = \me^{2C/3}\ast_8(H^{(2,2)}\wedge H^{(2,2)}) \,.
\end{equation}
This is the rotating version of the master equation \cite{Kim:2005ez,Kim:2006qu}. It reduces to the familiar non-rotating master equation of \cite{Kim:2006qu} by setting $e^{2C}=1$, $H^{(2,2)}=0$ and $\dd J = 0$ (so that $R_C = R_8$).

One can be slightly more explicit with the form of the flux terms and determine them up to primitive pieces. From \eqref{eq:da} and by decomposing $\dd a$ in term of its Hodge type we find 
\begin{equation}\label{eq:dapieces}
\begin{aligned}
\dd a^{(0)} &= \tfrac{4}{5}\, r^{-1} \me^{3B+C/3}\,\dd A^{(0)} \,, \\
\dd a^{(2,0)} &= \tfrac{1}{2} e^1 \wedge ( A^{(1,0)}-\ii\partial \alpha) + r\, \dd A^{(2,0)}\, ,\\
\dd a^{(0,2)} &= \tfrac{1}{2} \bar{e}^1 \wedge ( A^{(0,1)}+\ii\bar{\partial} \alpha) + r\, \dd A^{(0,2)}\, ,\\
\dd a_0^{(1,1)} &= \dd r\wedge\eta\,\big(\alpha+\tfrac{4}{5}\,\me^{3B+C/3}\,\dd A^{(0)}\big) + r \big( \alpha\, \dd\eta + \tfrac{1}{5}\,\dd A^{(0)} J + \dd A^{(1,1)}_0 \big) \\
&\phantom{=}\,\,\, + \tfrac{1}{2}\dd r \wedge (A - \dd^c \alpha) +\tfrac{1}{2}r\eta\wedge (J\cdot A - \dd \alpha) \, .
\end{aligned}
\end{equation}
We can use these decompositions to reduce \eqref{eq:traceflux} which implies:
\begin{equation}
\begin{aligned}
e^{-2C/3}\, J\lrcorner\, H^{(1,1)} &= 2\alpha\,,\\
e^{-2C/3}\, J \lrcorner\, H^{(2,1)} &= \ii \partial \alpha +  A^{(1,0)}\,, \\
e^{-2C/3}\, J \lrcorner\, H^{(2,2)} - e^{-3B-C}\, H^{(1,1)} &= 2\dd A^{(1,1)} + 2 \alpha \dd \eta \,.
\end{aligned}
\end{equation}
Therefore we may rewrite the fluxes as 
\begin{equation}\label{Hfluxdecomposition}
\begin{aligned}
H^{(1,1)} &=\tfrac{1}{2} e^{2C/3} \alpha J +H^{(1,1)}_0~, \\
H^{(2,1)} &= \tfrac{1}{3} e^{2C/3} J \wedge (i \partial \alpha + A^{(1,0)})+H^{(2,1)}_0~, \\
H^{(2,2)} &= \tfrac{1}{2} J \wedge \big( e^{-3B-C/3} H^{(1,1)} + 2e^{2C/3}( \dd A^{(1,1)}+\alpha \dd \eta)\big)\\
&\ \ \ -\tfrac{1}{3} ( 2e^{2C/3} \dd A^{(0)}+ e^{-3B+C/3} \alpha) J^2 +H^{(2,2)}_0~,
\end{aligned}
\end{equation}
where $H^{(p,q)}_{0}$ denotes the primitive piece. In principle one could now substitute these expressions into the Bianchi identities and Maxwell equations however this is not particularly enlightening and so we refrain from presenting them here. Note that the primitive pieces are essential for satisfying the Bianchi identities.

\subsection{Imposing the SO(2,1) isometry}\label{Toricpart}

So far our analysis has been for general scalars $C$, $\alpha$ and one-form $A$. However, in order to construct the near-horizon of a black hole we need to impose that there is an SO$(2,1)$ isometry, which leads to constraints on these fields. In appendix \ref{app:NH} we have given the general metric for the near-horizon of a rotating black hole with a manifest AdS$_2$ factor over which the internal manifold is fibered and seen the constraints that this imposes on the geometry. In particular the fibration is governed by a vector of constants $k^i$ associated to each Killing vector of the internal manifold fibered over AdS$_2$. As we reviewed in the appendix the necessity for these parameters to be constant arises in order that there is an SO$(2,1)$ isometry. From the analysis of appendix \ref{app:NH} we find that the scalar and one-form take the form\footnote{We use the math literature notation such that $g(\partial_{\phi_i}, \,\cdot)$ is a one-form.}
\be\label{NH C value}
\alpha \eta+A=- k^{i} g(\partial_{\phi_i}, \,\cdot)\, , \qquad \quad \me^{-2 C}= 1+ |\alpha \eta + A|_{9}^{2}\, ,
\ee
where $\partial_{\phi_{i}}$ are the Killing vectors of the internal manifold and the metric $g_{ij}$ is the metric on $\dd s^{2}_{9}$, as defined in \eqref{ansatzmetric}, restricted to the angular coordinates. Denoting by
\be
\eta_{i}\equiv g(\partial_{\phi_i},\, \cdot)\, ,
\ee 
the dual one-form of the Killing vector $\partial_{\phi_{i}}$ using the metric on $\dd s^{2}_{9}$. Then the one-form $a$ is simply
\be
a\equiv r(\alpha \eta + A)= -r k^{i} \eta_{i}\, .
\ee

In the remainder of this section let us assume that the 8d base is K\"ahler since this will allow for more explicit expressions. In addition we will assume that the base is toric, with the 9d space $Y_9$ admitting a U$(1)^5$ action with Killing vectors $\partial_{\phi_{i}}$.\footnote{We need not require the full space to be toric for our arguments to hold, we merely do so for simplicity of exposition. An interesting case to consider, which requires a minor generalization, is to consider a Riemann surface embedded into $Y_9$ as $Y_9\equiv O(\vec{n})_{\Sigma_{g}} \times_{ U(1)^{4}} Y_{7}$ with $\vec{n}$ a four-vector of constant twist parameters which are the Chern numbers of the U$(1)$ bundle over the Riemann surface \cite{Gauntlett:2018dpc}.} We may write the one form $\eta$ as\footnote{We follow the toric geometry notational conventions of \cite{Gauntlett:2018dpc}.}
\be
\eta = 2 \sum_{i} w_{i} \dd \phi_{i}\, ,\label{wscalar}
\ee
where the $w_{i}$ are the moment map coordinates of the cone restricted to $Y_9$. Moreover the K\"ahler two-form on the base may be expanded as
\be
J= \sum_{i} \dd x_{i} \wedge \dd \phi_{i}\, ,\label{xscalar}
\ee
where $x_{i}$ are global functions on $Y_9$ since $b_{1}(Y_9)=0$ for a toric contact structure. Note that
\be
\partial_{\phi_{i}}\lrcorner\, J= - \dd x_{i}\, .
\ee
With this short (and very incomplete) review of toric geometry we may proceed with writing the scalars and one-form in terms of the global functions of the toric geometry defined above. It follows that 
\be
\alpha =- k^{i} \partial_{\phi_{i}}\lrcorner\,  \eta=-2 k^{i} w_{i}\, .\label{alphatoric}
\ee
Next consider $A$, we find the simple result
\begin{align}
A&=-\me^{-3B-C/3}k^{i} \dd^{c}x_{i}\, .\label{Atoric}
\end{align}
Finally we may evaluate \eqref{NH C value} which implies
\be\label{rewritten C eq}
\me^{-2 C} =1+ (k^i w_i)^{2} + \me^{-3 B-C/3} |k^{i}\dd^{c} x_{i}|^{2}_{8}\, ,
\ee
where $|\cdot|_{8}^{2}$ is the norm with respect to the K\"ahler metric. In principle one could try to solve this for the scalar $C$, however this is a sextic equation to solve. One could use \eqref{rewritten C eq} as defining the combination $\me^{-3B - C/3}$ which appears ubiquitously in the geometry. 

Note that this last comment only applies when the gauge field $A$ is non-zero. When it vanishes and the fibration is only along the R-symmetry direction, it turns out that $C$ is constant. To see this it is more insightful to use the parametrization employed in appendix \ref{app:NH} where the $z$-coordinate is assigned its own constant $k^{z}$, i.e. we do not use the basis $\partial_{\phi_{i}}$ used previously in this section. In this basis the Killing vectors are the four U$(1)$ isometries of the base and the R-symmetry vector $\partial_{z}$. It is then clear that for $A$ to vanish each of the four constants associated to the U$(1)$'s of the base must be zero. It follows from \eqref{alpha:appexplicit} that $\alpha$ is precisely the constant $-k^{z}$. Moreover $\me^{-2C}$ takes the constant value,
\be
\me^{-2 C}=1+ (k^{z})^2\, .
\ee
The natural interpretation of this subcase is that of the near-horizon of a non-rotating black hole equipped with an electric component for the graviphoton and possibly including magnetic charges for each of the gauge fields in the 4d theory.


\section{Action for the theory}\label{sec:action}
One of the essential ingredients for performing the extremization in \cite{Couzens:2018wnk} was the existence of an action which gave rise to the equations of motion of the theory. This action was derived in \cite{Gauntlett:2007ts} for the near-horizon geometry of static black holes and strings in M-theory and Type IIB respectively. As a first step towards performing the extremization in the rotating case we will construct the analogous rotating action. Thereafter we impose the supersymmetry constraints on this action and show that it reduces to a simple and familiar form. The action computes the entropy of these black holes.

\subsection{Non-supersymmetric action}

The simplest method for constructing an action for the 9d geometry is to reduce the 11d action using our ans\"atze. By construction the equations of motion of the resulting 9d action will match the ones obtained in the section \ref{sec:8d susy}. We start from the action of eleven-dimensional supergravity
\begin{equation}
S_{11} = \frac{1}{2\kappa_{11}^2}\int \, R_{11} \ast_{11}\! 1 - \frac{1}{2}\, \mathcal{G}_4 \wedge \ast_{11}\, \mathcal{G}_4 - \frac{1}{6}\, \mathcal{C}_3 \wedge \mathcal{G}_4 \wedge \mathcal{G}_4 \,.
\end{equation}
Here $\mathcal{C}_3$ is the three-form potential and $\mathcal{G}_4 = \dd \mathcal{C}_3$ is its field strength. Using the ans\"atze
\begin{equation}
\begin{aligned}
\dd s_{11}^2 &= -\Delta^2 (\dd t + a)^2 + \Delta^{-1} e^{2\phi}\, \dd s_{10}^2 \,, \\
\mathcal{G}_4 &= \Delta^{-1} e^0 \wedge f_3 + h_4 \,,
\end{aligned}
\end{equation}
we reduce this action to 10d. The Bianchi identity $\dd\mathcal{G}_4 = 0$ implies that
\begin{equation}
\dd f_3 = 0 \,, \qquad\quad \dd h_4 + \dd a \wedge f_3 = 0 \,.
\end{equation}
We write these field strengths in terms of their potentials as
\begin{equation}
f_3 = \dd c_2 \,, \qquad\qquad h_4 = \dd c_3 - \dd a \wedge c_2 \,.
\end{equation}
Now we can write $\mathcal{G}_4$ into the convenient form
\begin{equation}
\mathcal{G}_4 = - \,\dd \big[ (\dd t + a) \wedge c_2 \big] + \tilde{h}^{(2,2)} \,,
\end{equation}
where we introduce the shifted four-form field strength
\begin{equation}
\tilde{h}^{(2,2)} = h_4 + \dd a \wedge c_2 = \dd c_3 \,.
\end{equation}
Note that although we add a superscript to indicate that upon imposing supersymmetry this field strength is a $(2,2)$-form, at the moment we have not imposed supersymmetry yet so we have to treat $\tilde{h}^{(2,2)}$ as a general four-form. The 11d potential $\mathcal{C}_3$ can now be expressed in terms of the 10d potentials as
\begin{equation}\label{decomp11dpotential}
\mathcal{C}_3 = -(\dd t + a) \wedge c_2 + {c}_3 \,.
\end{equation}
By using these ans\"atze and definitions, we find the 10d Lagrangian
\begin{equation}
\begin{aligned}
\mathscr{L}_{10} = \,&\, \Delta^{-3} e^{8\phi} \big( R_{10} - 72 (\partial_\mu \phi)^2 - 18\,\nabla^2\phi - 12(\partial_\mu \log \Delta)^2 + 7\,\nabla^2 \log\Delta + 56\, \partial_\mu \phi\, \partial^\mu \log \Delta \big) \ast_{10} 1 \\[4pt]
& + \tfrac{1}{2}\,e^{6\phi}\,\dd a\wedge\ast_{10}\,\dd a + \tfrac{1}{2}\,\Delta^{-3}e^{4\phi}\, f_3 \wedge \ast_{10}\, f_3 - \tfrac{1}{2}\, e^{2\phi}\, \tilde{h}^{(2,2)} \wedge \ast_{10}\, \tilde{h}^{(2,2)} \\[6pt]
& +e^{2\phi}\, c_2 \wedge \dd a \wedge \ast_{10}\,\tilde{h}^{(2,2)} -\tfrac{1}{2}\, e^{2\phi}\, c_2 \wedge \dd a \wedge \ast_{10}\, (c_2\wedge \dd a) \\[6pt]
& +\tfrac{1}{2}\, c_2 \wedge \tilde{h}^{(2,2)}\wedge \tilde{h}^{(2,2)} - \tfrac{1}{2} \, ({c}_2)^2 \wedge \dd a \wedge \tilde{h}^{(2,2)} + \tfrac{1}{6} (c_2)^3 \wedge (\dd a) ^2 \,.
\end{aligned}
\end{equation}
Next we want to consider the reduction of this Lagrangian to 9d, by using the cone ansatz presented in \eqref{sec:ansatz}. In addition, we want to split off the $\eta$-direction from the 8d space $\mathcal{B}$ so that we end up with a 9d Lagrangian density of the form $\mathscr{L}_9 = \eta\wedge(\ldots)$ where the dots represent an expression in terms of fields defined on $\mathcal{B}$. The relevant ans\"atze for this reduction are\footnote{Note that we omitted the part of $c_2$ that has one leg on $\dd r$ and one leg on $\mathcal{B}$ only. The reason for this is that such a term can be absorbed in $C_2$ by a gauge transformation.}
\begin{align}
\dd s^2_{10} &= \dd r^2 + r^2 \eta^2 + r^2 \,\me^{-3B-C/3}\, \dd s_8^2 \,,\nonumber\\
\me^{2\phi} &= r^{-3}\, \me^{3B+C} \,,\nonumber\\
\Delta &= r^{-1}\, \me^{B+C} \,,\nonumber\\
\dd a &= r\,(\alpha\,\dd\eta + \dd A) + \dd r \wedge A - r\,\eta\wedge \dd\alpha + \alpha\, \dd r \wedge \eta \,,\\
c_2 &= r^{-1} \,C_2 + r^{-1}\,\eta\wedge C_1 + r^{-2}\,C_0\,\dd r \wedge \eta \,,\nonumber\\
f_3 &= r^{-1} \,F_3 + r^{-2}\, \dd r \wedge F_2 + r^{-1}\, \eta \wedge \hat{F}_2 + r^{-2}\,\dd r \wedge \eta \wedge F_1 \,,\nonumber\\
\tilde{h}^{(2,2)} &= H^{(2,2)} + r^{-1} \dd r \wedge (H^{(2,1)} + H^{(1,2)}) + i\eta \wedge (H^{(2,1)} - H^{(1,2)}) + r^{-1} \dd r \wedge \eta \wedge H^{(1,1)} \,.\nonumber
\end{align}
Performing this reduction is a lengthy but in principle straightforward calculation. We find the 9d Lagrangian\footnote{We split off the $r$-coordinate as $\mathscr{L}_{10} = \mathscr{L}_9 \wedge r^{-2}\dd r$. Splitting off $\dd r$ on the left side would give an overall minus sign.}
\begin{equation}\label{eq:nonsusyaction}
\begin{aligned}
\mathscr{L}_{9} =\; &\, \eta\wedge \Big[ \big( R_{8} - \tfrac{9}{2} (\partial_\mu B)^2 -\tfrac{7}{6}(\partial_\mu C)^2 -3\, \partial_\mu B\, \partial^\mu C - 2\, e^{-3B-C/3} \big) \ast_{8} 1 -\tfrac{1}{2}\,e^{3B+C/3}\, \dd \eta \wedge\ast_8 \dd \eta \\
& +\tfrac{1}{2}\,e^{-3B+5C/3} \alpha^2 \ast_8 1 + \tfrac{1}{2}\,e^{2C} A\wedge\ast_8 A + \tfrac{1}{2}\,e^{2C} \dd \alpha \wedge\ast_8 \dd\alpha + \tfrac{1}{2}\,e^{3B+7C/3}(\alpha\,\dd\eta + \dd A)\wedge\ast_8(\alpha\,\dd\eta + \dd A) \\
& +\tfrac{1}{2}\,e^{-6B-2C} F_1\wedge\ast_8 F_1 + \tfrac{1}{2}\,e^{-3B-5C/3} F_2\wedge\ast_8 F_2 + \tfrac{1}{2}\,e^{-3B-5C/3} \hat{F}_2\wedge\ast_8 \hat{F}_2 + \tfrac{1}{2}\,e^{-4C/3} F_3\wedge\ast_8 F_3 \\
& -\tfrac{1}{2}\,e^{3B+C}\,H^{(2,2)} \wedge\ast_8 H^{(2,2)} -2e^{2C/3} H^{(2,1)} \wedge\ast_8 H^{(1,2)} -\tfrac{1}{2}\, e^{-3B+C/3} H^{(1,1)} \wedge\ast_8 H^{(1,1)} \\
& +e^{3B+C}\,C_2\wedge(\alpha\,\dd\eta+\dd A)\wedge\ast_8 H^{(2,2)} +ie^{2C/3} \big(C_1\wedge(\alpha\,\dd\eta+\dd A)-C_2 \wedge \dd\alpha\big) \wedge\ast_8 (H^{(2,1)}-H^{(1,2)}) \\
& + e^{2C/3} C_2\wedge A\wedge\ast_8 (H^{(2,1)}+H^{(1,2)}) +e^{-3B+C/3}\big(\alpha\, C_2 + C_1\wedge A + C_0\,(\alpha\,\dd\eta+\dd A)\big) \wedge\ast_8 H^{(1,1)} \\
& -\tfrac{1}{2}\,e^{3B+C}\,C_2\wedge(\alpha\,\dd\eta+\dd A)\wedge\ast_8 (C_2\wedge(\alpha\,\dd\eta+\dd A)) -\tfrac{1}{2}\, e^{2C/3} C_2\wedge A \wedge\ast_8 (C_2\wedge A) \\
& -\tfrac{1}{2}\,e^{2C/3} \big(C_1\wedge(\alpha\,\dd\eta+\dd A)-C_2\wedge \dd\alpha\big)\wedge\ast_8 \big(C_1\wedge(\alpha\,\dd\eta+\dd A)-C_2\wedge \dd\alpha\big) \\
& -\tfrac{1}{2}\,e^{-3B+C/3} \big(\alpha\, C_2 + C_1\wedge A + C_0\,(\alpha\,\dd\eta+\dd A)\big)\wedge\ast_8 \big(\alpha\, C_2 + C_1\wedge A + C_0\,(\alpha\,\dd\eta+\dd A)\big) \\
& -C_2\wedge H^{(2,2)}\wedge H^{(1,1)} -2iC_2\wedge H^{(2,1)}\wedge H^{(1,2)} -\tfrac{1}{2}\,C_0 \, H^{(2,2)}\wedge H^{(2,2)} -C_1\wedge H^{(2,2)}\wedge(H^{(2,1)} + H^{(1,2)}) \\
& +\tfrac{1}{2}\,(C_2)^2 \wedge \alpha\,H^{(2,2)} - \tfrac{1}{2}\,(C_2)^2\wedge\dd \alpha \wedge(H^{(2,1)}+H^{(1,2)}) - \tfrac{1}{2}\,i(C_2)^2\wedge A\wedge (H^{(2,1)}-H^{(1,2)}) \\
& +\tfrac{1}{2}\,(C_2)^2\wedge(\alpha\,\dd\eta+\dd A)\wedge H^{(1,1)} + C_0\,C_2\wedge(\alpha\,\dd\eta+\dd A)\wedge H^{(2,2)} +C_2\wedge C_1\wedge A\wedge H^{(2,2)} \\
& +C_2\wedge C_1\wedge (\alpha\,\dd\eta+\dd A) \wedge (H^{(2,1)}+H^{(1,2)}) -(C_2)^2\wedge C_1\wedge A\wedge(\alpha\,\dd\eta+\dd A) \\
& -\tfrac{1}{3} (C_2)^3\wedge\alpha\,(\alpha\,\dd\eta+\dd A) - \tfrac{1}{3}(C_2)^3\wedge A\wedge\dd \alpha - \tfrac{1}{2}\,C_0\,(C_2)^2\wedge(\alpha\,\dd\eta+\dd A)^2 \Big] \,.
\end{aligned}
\end{equation}
From this action one can derive the equation of motions that define the solutions discussed in the previous section \ref{sec:8d susy}. Note that we have not imposed any supersymmetry in deriving this action.

\subsection{Supersymmetric action}\label{ssec:susyaction}
Here we consider the restriction of the Lagrangian obtained above to off-shell supersymmetric geometries. We say these 9d geometries are off-shell because we do not impose the equations of motion such as \eqref{eq:box}, and supersymmetric since we do impose the supersymmetry constraints discussed in section \ref{sec:setup}. We will see that the Lagrangian \eqref{eq:nonsusyaction} becomes quite simple once supersymmetry has been imposed. The simplest method is to impose supersymmetry in 10d and subsequently reduce to 9d, instead of starting from the 9d Lagrangian \eqref{eq:nonsusyaction}. We begin with the 10d non-supersymmetric Lagrangian
\begin{equation}
\begin{aligned}
\mathscr{L}_{10} = \,&\, \Delta^{-3} \me^{8\phi} \big( R_{10} - 72 (\partial_\mu \phi)^2 - 18\,\nabla^2\phi - 12(\partial_\mu \log \Delta)^2 + 7\,\nabla^2 \log\Delta + 56\, \partial_\mu \phi\, \partial^\mu \log \Delta \big) \ast_{10} 1 \\[4pt]
& + \tfrac{1}{2}\,\me^{6\phi}\,\dd a\wedge\ast_{10}\,\dd a + \tfrac{1}{2}\,\Delta^{-3}\me^{4\phi}\, f_3 \wedge \ast_{10}\, f_3 - \tfrac{1}{2}\, \me^{2\phi}\, \tilde{h}^{(2,2)} \wedge \ast_{10}\, \tilde{h}^{(2,2)} \\[6pt]
& +\me^{2\phi}\, c_2 \wedge \dd a \wedge \ast_{10}\,\tilde{h}^{(2,2)} -\tfrac{1}{2}\, \me^{2\phi}\, c_2 \wedge \dd a \wedge \ast_{10}\, (c_2\wedge \dd a) \\[6pt]
& +\tfrac{1}{2}\, c_2 \wedge \tilde{h}^{(2,2)}\wedge \tilde{h}^{(2,2)} - \tfrac{1}{2} \, ({c}_2)^2 \wedge \dd a \wedge \tilde{h}^{(2,2)} + \tfrac{1}{6} (c_2)^3 \wedge (\dd a) ^2 \,.
\end{aligned}
\end{equation}
Here we can readily plug in the susy conditions
\begin{equation}
c_2 = \me^{2\phi}\,j \,, \qquad\quad f_3 = \dd(\me^{2\phi}\,j) \,.
\end{equation}
Furthermore, we use the decompositions
\begin{align}
\dd a &= \dd a^{(0)} \, j +\dd a_0^{(1,1)}+\dd a^{(2,0)}+\dd a^{(0,2)} \,,\\[4pt]
\tilde{h}^{(2,2)} &= \tilde{h}^{(2,2)}_0 + \tfrac{1}{2}\,\me^{2\phi}\,\dd a^{(0)}\, \frac{j^2}{2!} + \tfrac{2}{3}\,\me^{2\phi}\,\dd a^{(1,1)}_0 \wedge j \,,
\end{align}
to write out the Hodge stars
\begin{align}
\ast_{10}\,\dd a &= \dd a^{(0)} \, \frac{j^4}{4!} -\dd a_0^{(1,1)} \wedge \frac{j^3}{3!}+\big(\dd a^{(2,0)} + \dd a^{(0,2)}\big)\wedge \frac{j^3}{3!} \,, \\[4pt]
\ast_{10}\, (j \wedge \dd a) &= 2\,\dd a^{(0)} \, \frac{j^3}{3!} -\dd a_0^{(1,1)} \wedge \frac{j^2}{2!}+\big(\dd a^{(2,0)} + \dd a^{(0,2)}\big)\wedge \frac{j^2}{2!} \,,\\[4pt]
\ast_{10}\, \tilde{h}^{(2,2)} &= \tilde{h}^{(2,2)} \wedge j - \me^{2\phi}\,\dd a^{(0)} \, \frac{j^3}{3!} -2\,\me^{2\phi}\,\dd a_0^{(1,1)} \wedge \frac{j^2}{2!} \,.
\end{align}
By combining all these results, we find the 10d supersymmetric Lagrangian
\begin{equation}
\begin{aligned}
\mathscr{L}_{10}^{\,\text{SUSY}} = \,&\, \Delta^{-3} \me^{8\phi} \big( R_{10} - 80 (\partial_\mu \phi)^2 - 12(\partial_\mu \log \Delta)^2 + 62\, \partial_\mu \phi\, \partial^\mu \log \Delta \\[4pt]
& - \nabla^2(18\,\phi - 7\,\log\Delta) \big) \ast_{10} 1 + \tfrac{1}{2}\,\Delta^{-3}\me^{8\phi}\, \dd j \wedge \ast_{10}\, \dd j \,.
\end{aligned}
\end{equation}
Here we also used that $w_4 = j\lrcorner\,\dd j = 3\,\dd\log\Delta - 8\,\dd\phi$.

We reduce this Lagrangian to 9d using the ans\"atze
\begin{equation}
\begin{aligned}
\dd s^2_{10} &= \dd r^2 + r^2 \eta^2 + r^2 \,\me^{-3B-C/3}\, \dd s_8^2 \,,\\
\me^{2\phi} &= r^{-3}\, \me^{3B+C} \,,\\
\Delta &= r^{-1}\, \me^{B+C} \,,\\
j &= r\,\eta\wedge\dd r + r^2\,e^{-3B-C/3}\, J \,,
\end{aligned}
\end{equation}
and find (again using $\mathscr{L}_{10} = \mathscr{L}_9 \wedge r^{-2}\dd r$)
\begin{equation}
\begin{aligned}
\mathscr{L}_{9}^{\,\text{SUSY}} =\; &\, \eta\wedge \Big[ \big( R_{8} -\tfrac{3}{2}(\partial_\mu(3B+\tfrac{1}{3}C))^2 - 11\,\me^{-3B-C/3} \big) \ast_{8} 1 + 2\,J\wedge\ast_8\,\dd\eta \\[4pt]
& +2\,\me^{-3B-C/3}\,J\wedge\ast_8\, J + \tfrac{1}{2}\,\me^{6B+2C/3}\,\dd(\me^{-3B-C/3}\,J)\wedge\ast_8\,\dd(\me^{-3B-C/3}\,J) \Big] \,.
\end{aligned}
\end{equation}
We simplify this expression using the supersymmetry conditions
\begin{equation}
\begin{aligned}
J\lrcorner\,\dd\eta &= \me^{-3B-C/3} \,,\\
W_4 = J\lrcorner\,\dd J &= 0 \,,\\
R_C &= 2\,\me^{-3B-C/3} - \tfrac{2}{3}\,\Box C \,,
\end{aligned}
\end{equation}
as well as the relation between the Ricci and the Chern--Ricci scalar
\begin{equation}
R_C = R_8 +\tfrac{1}{2}\,|\dd J|^2 \,.
\end{equation}
This yields the surprisingly simple result
\begin{equation}
\begin{aligned}
\mathscr{L}_{9}^{\,\text{SUSY}} =\; &\, \eta\wedge \me^{-3B-C/3} \ast_{8} 1 \\[4pt]
=\; &\, \eta \wedge\dd\eta \wedge \frac{J^3}{3!} \,.\label{LSUSYNICE}
\end{aligned}
\end{equation}
Note that this is the same expression for the 9d supersymmetric action as was obtained in the non-rotating case in \cite{Couzens:2018wnk}. A subtle difference is that $\dd \eta \neq \rho$ here, but rather $\rho=\dd \eta+\frac{1}{3} \dd \dd^{c} C$. However since the forms $\rho$ and $\dd \eta$ are in the same cohomology class this distinction does not matter. Observe that
\begin{equation}
\begin{aligned}
\int_{Y_9} \eta \wedge \dd \dd^c C\wedge \frac{ J^3}{3!}&=\int_{Y_9} \eta \wedge \Bigg( \dd \Big( \dd^c C \wedge \frac{ J^3}{3!} \Big) + \dd^c C \wedge \frac{\dd J^3}{3!} \Bigg)=0\, ,
\end{aligned}
\end{equation}
where the first term equality uses the fact that $J, \dd J$ and $\dd^c C$ are basic\footnote{A form $\beta$ is basic with respect to $\xi$ if it satisfies both $\xi \lrcorner \beta=0$ and $\mathcal{L}_{\xi} \beta=0$. } with respect to to the R-symmetry vector $\xi$, and the second equality follows since the first term is a total derivative and the second vanishes because $J$ is balanced. We conclude that we may replace $\dd \eta$ by $\rho$ in expression \eqref{LSUSYNICE} and therefore the integrals for computing the supersymmetric action, and therefore the entropy
\be
S_{\mathrm{BH}}=\frac{1}{4 G_{11}} \int_{Y_9} \mathscr{L}_{9}^{\mathrm{SUSY}}\, ,\label{SBHSUSY}
\ee
in both the rotating and non-rotating cases are exactly the same.

Later in section \ref{sec:BlackString} we will discuss how one can obtain near-horizon geometries of rotating black strings in Type IIB from the 11d setup considered so far. In anticipation of this, let us reduce the 9d action for geometries on the M-theory side to a 7d action for geometries on the Type IIB side. These 7d geometries can be obtained from the 9d geometries by requiring that the 9d geometry admits a two-torus. By using the ansatz \eqref{decompJ6} we find the supersymmetric Lagrangian for the 7d geometry to be
\begin{equation}
\mathscr{L}_{7}^{\,\text{SUSY}} = \eta \wedge\dd\eta \wedge \frac{J_{(6)}^2}{2!} \,.
\end{equation}
Let us point out that one can replace $\dd \eta$ by $\rho_{(6)}$ in the 7d Lagrangian only when $\tau$ is constant. Namely, for a non-trivial axio-dilaton profile the term $\dd Q$ appearing in \eqref{eq:rho6} is only locally exact, and therefore cannot be interpreted as a total derivative term as was the case for the $\dd \dd^{c}C$ term. As studied in \cite{vanBeest:2020vlv} it is more convenient to view these near-horizon geometries from an 11d perspective rather than a 10d one. The central charge of the dual 2d SCFT is given by 
\be
c= \frac{3}{(2\pi)^6 \gs^2 \ls^8} \int_{Y_7} \eta \wedge \dd \eta \wedge \frac{J_{(6)}^2}{2!}\label{SBSIIB}\, .
\ee


\section{Embedding of the AdS$_\text{4}$ Kerr--Newman black hole}\label{Sec:Example}

Here we study the embedding of the supersymmetric limit of the AdS$_4$ Kerr--Newman (KN) black hole solution found in \cite{PhysRev.174.1559} and further studied in \cite{Kostelecky:1995ei,Caldarelli:1998hg} into our classificatio by using the uplift of minimal gauged supergravity on an arbitrary 7d Sasaki--Einstein manifold. Note that we could have taken one out of the zoo of supersymmetric rotating AdS$_4$ solutions, e.g. \cite{Hristov:2019mqp,Hosseini:2020mut,Cvetic:2005zi,Chow:2013gba}. We  choose the KN solution since it is the simplest yet contains all the necessary ingredients. We begin by considering the black hole in four dimensions before studying the full eleven-dimensional solution.


\subsection{Kerr--Newman solution}

The four-dimensional black hole is given by
\begin{align}
\dd s^2&= -\frac{\Delta_r}{W}\bigg( \dd t -\frac{\bc \sin^2 \theta}{\Xi} \dd \phi\bigg)^2 +W \bigg( \frac{\dd r^2}{\Delta_r}+\frac{\dd \theta^2}{\Delta_{\theta}} \bigg)+ \frac{\Delta_{\theta} \sin^2 \theta}{W} \bigg( \bc \dd t-\frac{\tilde{r}^2+\bc^2}{\Xi} \dd \phi \bigg)^2\, ,\\
A&= \frac{2 m \tilde{r} \sinh^2 \delta}{W}\bigg( \dd t -\frac{\bc \sin^2 \theta}{\Xi} \dd \phi \bigg) +\alpha_{\text{gauge}} \dd t\, ,
\end{align}
where 
\begin{align}\label{AdS4BHSolutionFunctions}
\tilde{r}&=r+2m \sinh^2 \delta\, ,\nonumber\\
\Delta_r&= r^2 +\bc^2 -2 m r + \tilde{r}^2(\tilde{r}^2 +\bc^2)\, ,\nonumber\\
\Delta_\theta&=1- \bc^2 \cos^2 \theta\, ,\\
W&= \tilde{r}^2 + \bc^2 \cos^2 \theta\, ,\nonumber\\
\Xi&=1-\bc^2\, . \nonumber
\end{align}
The solution is characterised by three constants $(\bc, \delta, m)$ whilst the parameter $\alpha_{\text{gauge}}$ is related to a pure gauge transformation and is therefore not a parameter of the solution. The solution describes a non-extremal black hole provided that $\bc^2<1$ and $m$ is bounded from below. The exact value of the bound is not important for our purposes, but it is derived in \cite{Caldarelli:1999xj}. Without loss of generality we have $m,\delta,\bc>0$. The black hole is characterized by its energy $E$, electric charge $Q$ and momentum $J$:
\begin{equation}
E = \frac{m}{G_{(4)}\Xi^2} \cosh 2\delta \, , \qquad Q=\frac{m}{G_{(4)}\Xi} \sinh 2 \delta \, , \qquad J=\frac{m\bc}{G_{(4)}\Xi^2} \cosh 2 \delta \, .
\end{equation}
The Bekenstein-Hawking entropy of the black hole can be found by computing the area of the outer horizon, resulting in
\begin{equation}
S =\left. \frac{\pi(\tilde{r}^2+\bc^2)}{G_{(4)}\Xi}\right|_{r=r_+} \, ,\label{KNentropy}
\end{equation}
where $r_+$ denotes the largest positive root of $\Delta_r=0$, and therefore describes the location of the outer horizon. For arbitrary values of the parameters $(\bc, \delta, m)$, the black hole is neither extremal nor supersymmetric. The BPS limit is defined by first imposing supersymmetry and then extremality. The supersymmetry is attained by imposing
\be
\me^{4 \delta}=1+ \frac{2}{\bc}\, .
\ee 
The solution is now supersymmetric but not extremal, in fact it has timelike closed curves and a naked singularity. To remedy this and obtain an extremal black hole we further identify
\be
m=\bc(1+\bc) \sqrt{2+\bc}\, .
\ee
There is now only a single parameter left in the theory, namely $\bc$. With these identifications the function $\Delta_r$ acquires a double root at 
\be
r^{*}=\bc\sqrt{2+\bc}\left(1+\bc-\sqrt{\bc(2+\bc)}\right) \, ,
\ee
with the other two roots becoming complex. 


\subsection{Near-horizon limit} 
We now want to take the near-horizon limit of the solution. It is convenient through a change of coordinates to shift the double root location in $\Delta_r$ to $0$ and to rewrite the function as
\be
\Delta_r= \rho^2 f(\rho)\, ,\quad f(\rho)=(\rho+r^*-r^-)(\rho+r^*-r^+) \, , 
\ee
where 
\be
\rho=r- r^*\, .
\ee
Since we will need to evaluate the function $f$ at the horizon often, we note that
\be
f(0) = 1+\bc\,(6+\bc) \, .
\ee
In the metric, the change of the $r$ to $\rho$ coordinate results only in changes in the functions \eqref{AdS4BHSolutionFunctions}, since the $\dd r$ term is invariant. To simplify notation we will therefore shift the functions such that an argument of $0$ means we evaluate at the horizon. In particular we now take
\be
\tilde{r}(\rho) = \rho + r^* + 2m \sinh^2 \delta \ ,
\ee
such that $\tilde{r}(0)$ is evaluating the function $\tilde{r}$ at the horizon. Similarly $W(0,\theta)$ evaluates $W$ at the horizon; for notational convenience we denote the functions $W(0,\theta)=W(\theta)$ and $f(0)=f_0$. Furthermore, in the BPS limit one can derive that $\tilde{r}(0) = \sqrt{\bc}$.
To take the near-horizon we perform the change of coordinates
\be
\rho \rightarrow \epsilon \rho\, ,~~~t\rightarrow \frac{t}{\epsilon} \, ,~~~\phi\rightarrow \phi+  \frac{\beta t}{\epsilon} \, ,
\ee
where $\beta$ is a constant that we will determine shortly an then send $\epsilon\rightarrow 0$. The near-horizon limit is now obtained by taking $\epsilon\rightarrow 0$ after making the above substitutions. The $\dd \theta^2$ term will clearly be sent to $W(\theta)/\Delta_{\theta}$, and we can ignore this term for time being. We find
\begin{align}
\frac{\Delta_r}{W(r,\theta)}\bigg( \dd t -\frac{\bc \sin^2 \theta}{\Xi} \dd \phi\bigg)^2&\rightarrow \frac{\rho^2f_0}{W(\theta)} \bigg(1-\frac{\beta \bc \sin^2 \theta}{\Xi}\bigg)^2 \dd t^2\, ,\nonumber\\[5pt] 
\frac{W(r,\theta)}{\Delta_r} \dd r^2&\rightarrow  \frac{W(\theta)}{f_0}\frac{\dd \rho^2}{\rho^2}\, ,\\[5pt]
\frac{\Delta_{\theta} \sin^2 \theta}{W(r,\theta)} \bigg( \bc \dd t-\frac{\tilde{r}^2+\bc^2}{\Xi} \dd \phi \bigg)^2&\rightarrow \frac{\Delta_\theta \sin^2 \theta}{W(\theta)}\bigg[ \frac{\dd t}{\epsilon}\Big(\bc- \frac{\tilde{r}(\epsilon \rho)^2+\bc^2}{\Xi} \beta\Big)- \frac{\bc+\bc^2}{\Xi} \dd \phi\bigg]^2\, .\nonumber
\end{align}
In the last line we can expand $\tilde{r}(\epsilon \rho)\sim \sqrt{\bc}+ \epsilon \rho \tilde{r}'(0)+\mathcal{O}(\epsilon^2)$, resulting in a term which diverges as $\epsilon^{-1}$, proportional to the constant
\be
1- \frac{1+\bc}{\Xi}\beta \, .
\ee
The existence of this term is the reason we introduced the shift in the $\phi$ coordinate, and it can be set to zero by fixing the constant $\beta$ in the shift as
\be
\beta = \frac{ \Xi}{1+ \bc}\, .
\ee
Including this factor of $\beta$ we can combine the results from above and write down the final result for the near-horizon solution
\begin{align}\label{NearHorizonMetricKlemm}
\begin{split}
\dd s^2|_{\mathrm{NH}}&=\frac{W(\theta)}{f_0}\bigg(- \rho^2 \dd t^2+ \frac{\dd \rho^2}{\rho^2}\bigg)\\
&\quad+\frac{W(\theta)}{\Delta_{\theta}} \dd \theta^2+\frac{\sin^2 \theta \, \Delta_{\theta}}{W(\theta)} \Big(\frac{\bc+\bc^2}{\Xi}\Big)^2\bigg(\dd \phi+ \frac{2 \sqrt{\bc} \, \Xi }{(1+\bc)f_0}\, \rho \, \dd t\bigg)^2\, ,
\end{split}
\end{align}
where, in order to make the AdS$_2$ factor manifest, we rescaled the time-coordinate
\be\label{RescaledTimeCoordinate}
t \rightarrow \frac{\bc(1+\bc)}{f_0} t \, .
\ee
Consider now the gauge field. Performing the same near-horizon limit and imposing the BPS limit, we find a divergent term in the gauge field, proportional to
\be
\frac{dt}{\epsilon}(\alpha_{\text{gauge}}-2) \, .
\ee
This term is purely gauge and we can remove it without problem by making a suitable choice for the gauge parameter. The resulting near-horizon vector field is
\be
\left. A\right|_{\mathrm{NH}}=\frac{2 \, \sqrt{\bc} \, (1+\bc)}{W(\theta)}\bigg( \frac{2\bc-W(\theta)}{f_0} \rho \, \dd t+\frac{\bc \sqrt{\bc} \sin^2 \theta}{\Xi} \dd \phi\bigg)\, ,
\ee
where of course the time coordinate has been rescaled with the same factor \eqref{RescaledTimeCoordinate} as in the metric.

\subsection{Uplift to 11d}
Now that we have derived the near-horizon metric and gauge field of the AdS$_4$ KN solution in minimal supergravity we can consider the uplift to 11d supergravity. The uplift of the metric and flux to eleven dimensions are given by
\begin{align}\label{UpliftFormula}
\begin{split}
\dd s_{11}^2 &= \dd s_4^2+ \left(\eta+ \tfrac{1}{4} A \right)^2  + \dd s_6^2 \, , \\
\mathcal{G}_4 &= \frac{3}{8}\, \text{dvol}(\text{AdS}_4) -\frac{1}{4} \star_4 F \wedge J\, ,
\end{split}
\end{align}
where $F=dA$ is the field strength of $A$, $\dd s_4^{2}$ is the near-horizon metric we just derived in \eqref{NearHorizonMetricKlemm} and $\dd s_6^{2}$ is the base of the Sasaki-Einstein manifold with $\eta = \dd z + \sigma$ dual to the Reeb-vector $\partial_{z}$. The conventions are chosen such that $\dd \eta = 2 J$, where $J$ is the K\"{a}hler form on $\dd s_6^2$. 

We now want to rewrite the metric and flux appearing in \eqref{UpliftFormula} in the form of our classification as presented in section \ref{sec:8d susy}. To recover this form, we write the metric in \eqref{UpliftFormula} such that it becomes a time-fibration over a base. It is also necessary to perform some coordinate redefinitions
\begin{align}\label{PhiPsiCoordinateShift}
\begin{split}
z &\rightarrow \frac{\bc\, (3+\bc)}{2 \,f_0}\,z \, , \\
\phi &\rightarrow  \phi - \frac{\Xi}{f_0}\,z \, .
\end{split}
\end{align}
After completing the straightforward but tedious rotations of the vielbeins and shifting the coordinates, the metric we find is of the following form
\be\label{UpliftedBobevMetric}
\dd s^2_{11}= \me^{2 B} \Big[ -\frac{\me^{2 C}}{r^2} (\dd t+ a)^2+ \frac{\dd r^2}{r^2}+ \eta^2 + Y(\theta)\, \dD\phi^2+\frac{f_0}{\Delta_{\theta}}\, \dd\theta^2+\me^{-2B}\dd s_6^2\Big]\, .
\ee
We will now clarify the several notational conventions used in this metric. Firstly, we have renamed the coordinate $\rho$ to $r$, in order to conform with the conventions of the classification. We have also introduced the function $Y$, $\dD\phi$ and redefined $\eta$ as
\begin{align}
\begin{split}
Y(\theta) &= \frac{\bc^2\,(3+\bc)^2 \, f_0 \, (1-\bc^2 \cos^2 \theta)\,\sin^2 \theta}{ \Xi^2 \cos^2 \theta \left(2\bc+W(\theta)\right)^2} \, , \\[5pt]
\dD \phi &= \dd \phi + \frac{2 \, \Xi}{\bc\,(3+\bc)}\, \sigma \, , \\[5pt]
\eta &= \dd z + \frac{2 \,f_0}{\bc\,(3+\bc)}\,\sigma + \frac{ f_0 \, (\bc \sin \theta)^2}{\Xi \,\left(2\bc+W(\theta)\right)} \,\dD \phi \, .
\end{split}
\end{align}
Note that the coordinate shift we made in \eqref{PhiPsiCoordinateShift} was necessary to ensure that the metric ends up with $\dd z^2$, with its coefficient being exactly equal to one. The scalars $\me^C$ and $\me^B$ are found to be
\begin{align}
\me^{C} &= \frac{\bc(1+\bc)\, \cos \theta \, \left(W(\theta)+2\bc\right)}{W(\theta)\sqrt{f_0\, W(\theta)}} \, , \\
\me^{B} &= \sqrt{\frac{W(\theta)}{4 \, f_0}} \, .
\end{align}
Recall that these scalars can also be used to compute $\Delta$ in \eqref{DeltaAndPhiScaling}.
The last remaining puzzle-piece in the metric is the fibration $a$, which is given by
\be\label{BobevCrichignoTimeFibration}
a = \sqrt{\bc} \,r \left( \frac{2 \,\bc^2 \cos^2 \theta - \bc^2 - 1}{(1+\bc)\, \left(W(\theta)+2\bc\right)} \, \eta + \frac{\bc \,(3+\bc)\,\tan^2 \theta \,(\bc^2 \, \cos^2 \theta - 1)f_0}{(1+\bc) \, \Xi   \left(W(\theta)+2 \bc\right)^2}\, \dD \phi \right) \, .
\ee
The fibration is of the expected form $a=r( \alpha \, \eta + A)$, and this specification of $a$ completes the endeavour of writing the metric in the classification form.
Now we can move on to consider the flux; recall that in the classification we wrote it as
\begin{align}
\begin{split}
\mathcal{G}_4 &= - \dd \left( (\dd t+ a) \wedge \me^{2\phi} j\right) + \tilde{h}^{(2,2)}\, , \\
\tilde{h}^{(2,2)} &= \dd c_3 \, .
\end{split}
\end{align}
We have already found the fibration $a$ in \eqref{BobevCrichignoTimeFibration} and $\me^{2\phi}$ is given in terms of the scalars $\me^B$ and $\me^C$, by making use of \eqref{DeltaAndPhiScaling}. The ten-dimensional complex structure form $j$ can be found from the vielbeins of the metric we found in \eqref{UpliftedBobevMetric}. Our remaining tasks thus consists of finding an expression for $\tilde{h}^{(2,2)}$, which in its turn is determined by the potential $c_3$. The only form we have thus not yet specified is the potential $c_3$. After carefully rewriting the flux we obtain from \eqref{UpliftFormula}, the resulting potential is given by
\begin{align}
\begin{split}
c_3 &= \frac{\bc^2 \sqrt{\bc} \,(3+\bc)}{8 \, \Xi \,\left(2\bc + W(\theta) \right)} \left(\sin \theta \,\dd\theta \wedge \eta \wedge \dD \phi  +\frac{(3+\bc)(2\bc-W(\theta))}{\bc \,\Xi\,\cos\theta} \, \dD \phi \wedge \dd(\dD \phi)   \right. \\[5pt]
&\quad  \left.+ \frac{\sin^2 \theta \left(2\bc-W(\theta)\right)}{r \, f_0\,\bc \cos \theta} \, \dd r \wedge \dD \phi \wedge \eta \right) \, .
\end{split}
\end{align}
From the above expression we can immediately determine that all components of the flux in \eqref{Hfluxin8Dcomponents} are turned on. Having obtained the potential, and thus its field strength, we have completed our mission of embedding the AdS$_4$ black hole solution into the classification. We have checked that the solution satisfies all the conditions of our classification, which is a non-trivial check of the correctness of our results.

Another useful way to write the metric consists of explicitly showing the AdS$_2$ factor we also obtained in \eqref{NearHorizonMetricKlemm}, which will allow us to read off the values of the near-horizon angular velocities, denoted by $k^i$. These $k^i$ are needed to define the killing vectors, over which we should integrate to find the angular momentum of the black hole. To obtain the particular form of the metric, we undo the rotation of the vielbeins (or simply do not rotate them in the first place). Instead of the time-fibration we then find a metric reminiscent of the one written in \eqref{AdS2NH}
\begin{align}\label{11DExplicitAdS2Metric}
\begin{split}
\dd s^2 &= \me^{2 B}\left( -r^2 \dd t^2 +\frac{\dd r^2}{r^2}\right) +\gamma_{\theta\theta}\, \dd \theta^2 +\dd s_6^2 \\[5pt]
&\quad+\gamma_{\mu\nu}\left( \dd \psi^{\mu}+M^{\mu}(\theta) \, \sigma +k^{\mu} r \dd t\right)\,\left( \dd \psi^{\nu}+M^{\nu}(\theta) \, \sigma+k^{\nu} r \dd t\right)  \, , 
\end{split}
\end{align}
where the AdS$_2$ is now clearly visible. As before, $\dd s_6$ denotes the base of the Sasaki-Einstein manifold and the one-form $\sigma$ is still defined on the K\"{a}hler-Einstein space $\dd s_6$ such that $\dd \sigma = 2 J$. Apart from these already familiar notions we established several new notational conventions; first of all we have introduced $\gamma_{\theta\theta}$ and $M^{\mu}(\theta)$ as
\begin{align}
\gamma_{\theta\theta} &=  \frac{W(\theta)}{4\, \Delta_{\theta}} \, ,\\
M^z &= \frac{2 \,f_0}{\bc\,(3+\bc)} \, ,\\
M^\phi &= \frac{2 \, \Xi}{\bc\,(3+\bc)} \, .
\end{align}
Besides these coefficients we introduced indices $\mu,\nu\in \{z, \phi \}$, along with a metric $\gamma_{\mu\nu}$ we will specify below and, finally, defined $\dd\psi$ as
\be
\dd\psi^\mu = (\dd z,\, \dd\phi) \, .
\ee
The metric \eqref{11DExplicitAdS2Metric} shows that only the $\phi$ and $z$ coordinates are gauged over the AdS$_2$ space. We could have expected this, since the original AdS$_4$ black hole had rotation only in the $\phi$ direction, and in \eqref{UpliftFormula} we have gauged the Reeb-vector with respect to the four-dimensional gauge vector. The metric, $\gamma_{\mu \nu}$,  we introduced for these two coordinates has the following components
\begin{align}
\gamma_{z z} &= \kappa^2 - \frac{\bc^2 \, (1+\bc) \sin^2 \theta}{f_0 \, W(\theta)} \left( \kappa - \frac{\bc \, (1+\bc)\, N(\theta)}{4\, f_0 \, W(\theta)} \right) \, , \\[5pt]
\gamma_{z \phi} &= \frac{\bc^2 \, (1+\bc) \, \sin^2 \theta}{ 2\,\Xi \, W(\theta)} \left(\kappa - \frac{\bc \, (1+ \bc) \, N(\theta) }{2 \, f_0 \, W(\theta)}    \right) \, , \\[5pt]
\gamma_{\phi \phi} &= \frac{\bc^3 \, (1+ \bc)^2 \, \sin^2 \theta \, N(\theta)}{4 \, \Xi^2 \, W(\theta)^2} \, ,
\end{align}
where, to alleviate the notational clutter, we have introduced the constant $\kappa$ and the function $N(\theta)$ as
\begin{align}
\begin{split}
\kappa &= \frac{\bc\, (3+\bc)}{2 \,f_0} \, , \\
N(\theta) &= 1+\bc-\bc^2 \cos^2 \theta -\bc^3 \, \cos^4 \theta \, .
\end{split}
\end{align}
Now that we have specified the $\gamma_{\mu \nu}$ in \eqref{11DExplicitAdS2Metric}, the description of the metric is almost complete. The last remaining unknowns are the constants $k^i$ which specify the gauging over the AdS$_2$. We find 
\begin{align}
k^z = k^\phi= \frac{1-\bc}{\sqrt{\bc}\, (3+\bc)} \, .\label{kiKN}
\end{align}
Note that the precise value of $k^{\phi}$ depends on how we scale the $\phi$ coordinate; the fact that both $k^i$ are equal arises due to our conventions for the coordinates. 
Since the above $k^i$ are the only non-zero ones, it follows that the AdS$_4$ black hole rotates only in the $z$ and $\phi$ directions. We can now check the identifications \eqref{alphatoric} and \eqref{Atoric}. In order to do this we need to compute the scalars $w_i$ and $x_{i}$ appearing in \eqref{wscalar} and \eqref{xscalar} for our solution. For simplicity we use the basis of Killing vectors $\{z, \phi, \psi_{1},\psi_{2},\psi_{3}\}$, with the $\psi_{1,2,3}$ the Killing vectors of the 6d K\"ahler-Einstein base of the Sasaki--Einstein space. In the following we need only compute the scalars for $z$ and $\phi$ since the solution only rotates in the $z$ and $\phi$ directions. We find
\begin{equation}
\begin{aligned}
x_{\phi}&=- \frac{\bc^{5/3}(3+\bc)(1+\bc)^{1/3}(3+\bc \cos^{2} \theta)^{1/3}}{8 \Xi  f_{0}^{2/3}}\, ,\quad\quad &x_{z}&=0\, ,\\
w_{\phi}&=\frac{\bc f_0 \sin^2 \theta}{2 \Xi (3+ \bc \cos^{2} \theta)}\, ,\quad \quad &w_{z}&=\frac{1}{2}\, .
\end{aligned}
\end{equation}
It is then a simple matter of substituting these and the constants $k^{i}$ found in \eqref{kiKN} to see that both \eqref{alphatoric} and \eqref{Atoric} are satisfied. Moreover using the results of section \ref{ssec:susyaction} and appendix \ref{app:NH} we can compute the entropy. With a little care in the definitions of the periods it follows that the integral \eqref{SBHSUSY} is the same as \eqref{KNentropy} and therefore this serves as another consistency check of our identification of the entropy and the supersymmetric action in \eqref{SBHSUSY}.


\section{Black strings in Type IIB}\label{sec:BlackString}

Having studied our 11d setup we now turn our attention to rotating black string solutions in Type IIB supergravity. We take our 11d setup and require that the internal space admits a two-torus, $T^2$. The 8d balanced manifold then breaks up as a semidirect product of this torus and a 6d manifold. Wherever 8d quantities split up in components on the torus and the 6d manifold, we simply denote this with the subscripts (2) and (6). Under this assumption of a torus in the internal space we can apply dualities to arrive in Type IIB, where we find a classification of rotating black string solutions that can be interpreted as rotating D3-branes wrapped on a Riemann surface.

If we add a warp factor acting homogeneously on the torus, the balanced condition of the 8d manifold implies that the 6d manifold is conformally balanced. For simplicity we do not take into account such a warping which gives a balanced 6d manifold. As such we take the metric ansatz
\be\label{metrictorusansatz}
\dd s_{11}^2=\me^{2 B} \Big[ -\me^{2C}\Big( \frac{\dd t}{r} + \alpha \eta + A_{(6)}+ A_{(2)}\Big)^2 +\frac{\dd r^2}{r^2}+ \eta^{2} + \me^{-3B-C/3}\Big(\dd s^{2}_{6}+ \frac{1}{\tau_{2}}(\dd x+ \tau_{1} \dd y)^2 + \tau_{2} \dd y^2\Big)\Big] \,,
\ee
where $\tau_{1}$ and $\tau_{2}$ are scalars valued on the 6d base and the complex combination $\tau=\tau_{1}+\ii \tau_{2}$ is a holomorphic function ($\bar{\partial}\tau = 0$). In principle we can take the two U$(1)$'s of the two-torus to be fibered over AdS$_2$, i.e. in the language of appendix \ref{app:NH} we can introduce constants $k^{x},k^{y}$ which are related to the angular momenta in these directions. However, introducing these parameters leads to the system becoming unreasonably complicated\footnote{In Type IIB these extra parameters will lead to a further warping of the metric. In particular, the dilaton will not be simply the dilaton one would get from the F-theory picture, i.e. $\tau_2^{-1}$. In addition, since we must satisfy \eqref{eq:traceflux} it is clear that turning these on will lead to turning on additional fluxes other than the self-dual five-form in Type IIB. It would be interesting to fully work out the details of this more general case, but it deserves more than this small section in this paper and a full treatment of the most general construction.} once we arrive in Type IIB, and therefore we shall just proceed with these parameters set to zero, which in \eqref{metrictorusansatz} implies that $A_{(2)} = 0$. In addition, we also assume that $\eta$ has no dependence on the $T^2$. The final piece of the solution we need to specify is the dependence of the flux on the torus: we take $\tilde{h}^{(2,2)}$ to have no legs along the torus directions.\footnote{The primitive piece of this part of flux (with legs on the torus) will give rise to a transgression term like in \cite{Donos:2008ug, Couzenstoappear}. Again for our purposes such a term is an unnecessary complication, and so we set it to zero here, although it is certainly interesting to consider.} Note that this is consistent with setting the rotation of the solution along the torus directions to zero, through the condition \eqref{eq:traceflux}. In addition to this, we assume that the scalars $B,C$ are independent of the torus coordinates, and are hence defined on the 6d base.

We now reduce the 8d conditions from section \ref{sec:8d susy} with this assumption of a torus in the internal space onto a set of conditions on the inherited 6d base space that has an SU$(3)$ structure. We decompose the two-form as
\be\label{decompJ6}
J= J_{(6)} +J_{(2)} \, ,\qquad\quad J_{(2)}=\dd x \wedge \dd y \, ,
\ee
which (using \eqref{eq:balance}) implies that $J_{(6)}$ is a balanced two-form: $\dd J_{(6)}^2 = 0$. Furthermore from \eqref{eq:tracedeta} we find that
\begin{equation}
\dd\eta\wedge\frac{J_{(6)}^2}{2!} = \me^{-3B-C/3} \, \frac{J_{(6)}^3}{3!} \,,
\end{equation}
which implies $J_{(6)}\lrcorner\,\dd\eta = \me^{-3B-C/3}$. We write the holomorphic four-form as
\be
\Omega= \Omega_{(6)} \wedge \Omega_{(2)}\, , \qquad\quad \Omega_{(2)}= \frac{1}{\sqrt{\tau_2}}\, \big(\dd x + \tau \dd y\big) \,.
\ee
From \eqref{eq:ricciform} it now follows that
\begin{equation}
\dd\Omega_{(6)} = \ii\big( P + \tfrac{1}{3}\dd^c C - Q \big) \wedge \Omega_{(6)} \,,
\end{equation}
where $Q= -\frac{1}{2 \tau_{2}} \dd \tau_{1}$ and we have used the holomorphicity of $\tau$. This gives us the Ricci form on the 6d space as
\be
\rho_{(6)}= \dd \eta +\tfrac{1}{3} \dd \dd^{c} C - \dd Q\, ,\label{eq:rho6}
\ee
which is the generalization of equation (2.57) of \cite{Couzens:2017nnr} to the rotating case. The additional term changes the expression for the Chern--Ricci scalar to
\be
R_{C(6)}= 2\me^{-3 B- C/3} -\tfrac{2}{3} \square_6 C+\frac{1}{2\tau^2_{2}} \,|\dd \tau|^2\, .
\ee
With our ansatz the 8d Bianchi identities \eqref{8dbianchis} for the fluxes $H^{(p,q)}$ remain the same but should be understood as 6d conditions. The expansions of these fluxes as in \eqref{Hfluxdecomposition} require slight modifications in the numerical coefficients but are otherwise the same after the replacement $J\rightarrow J_{(6)}$. From reducing the Maxwell equations (\ref{maxwell1} --\! \ref{maxwell4}) we find
\begin{equation}
\begin{aligned}
- \dd \star_{6}\dd \me^{-3 B-C}+ \me^{-2 C/3} \dd \eta \wedge \dd \eta \wedge J_{(6)}&=0\, ,\\
\dd \eta \wedge \dd (\me^{-2C/3} J_{(6)})&=0\, ,\label{MaxwellonT2}\\
\dd \dd^{c} (\me^{-2C/3} J_{(6)})&=0\, ,\\
\dd \star_{6} \dd \me^{-2 C/3}&= H^{(2,2)}\wedge H^{(1,1)} +2 \ii H^{(2,1)}\wedge H^{(1,2)}\, .
\end{aligned}
\end{equation}

We can now proceed by reducing along the A-cycle $(\dd x+ \tau_{1} \dd y)$ of the torus to Type IIA supergravity. Note that the Ricci form is independent of the $T^2$-coordinates and therefore so is the one-form $\eta$. This leads to a standard reduction of 11d supergravity to massless Type IIA. One finds that the metric in string frame is given by
\begin{align}
\dd s^2_{\text{IIA}}&=\me^{2B + 2\phi_{\text{IIA}}/3} \Big[ - \me^{2 C}\Big( \frac{\dd t}{r}+ \alpha \eta + A_{(6)}\Big)^{2} +\frac{\dd r^{2}}{r^2}+\eta^2+ \me^{-3B-C/3}\big( \dd s^{2}_{6}+ \tau_2 \dd y^2 \big) \Big]
\end{align}
and is supplemented by 
\begin{align}
\me^{4 \phi_{\text{IIA}}/3}&= \frac{1}{\tau_{2}}\me^{-B-C/3} \, ,\\
C_{1}^{\text{IIA}}&= \tau_{1}\dd y\, ,\\
C_{3}^{\text{IIA}}&= c_3 - \me^{3 B+C}\Big(\frac{\dd t}{r}+\alpha \eta + A_{(6)}\Big)\wedge \Big( \eta \wedge \frac{\dd r}{r}+\me^{-3B-C/3} J_{(6)}\Big)\, ,\\
B_2^{\text{IIA}}&=\me^{2C/3}\Big( \frac{\dd t}{r}+\alpha \eta + A_{(6)}\Big) \wedge \dd y\, .
\end{align}
Recall that we can decompose the 11d gauge potential as \eqref{decomp11dpotential}, where $c_3$ is the potential corresponding to $\tilde{h}^{(2,2)}$, and $c_2 = \me^{2\phi} j$ is fixed by supersymmetry.

By performing a T-duality along the $y$-direction we land in Type IIB. The metric in Einstein frame reads
\begin{align}
\dd s^{2}_{\text{IIB}}=\;&\me^{3 B/2 -C/6}\Big[ -\me^{2 C}\Big( \frac{\dd t}{r}+ \alpha \eta + A_{(6)}\Big)^2 +\me^{2C/3} \Big(\dd y + \me^{2 C/3}\Big( \frac{\dd t}{r} + \alpha \eta + A_{(6)}\Big)\Big)^2\nonumber\\[4pt]
&+\frac{1}{r^2}\Big(\dd r^{2}+r^2 \big(\eta^2+\me^{-3B-C/3} \,\dd s^{2}_{6}\big)\Big)\Big]\, .\label{IIBcone}
\end{align}
Here we have made explicit a cone in the geometry. It is useful to redefine the scalar $B$ in the form $B=-\widetilde{B}/3+C/9$ which puts the metric in the form
\begin{align}
\dd s^{2}_{\text{IIB}}=\;&\me^{- \widetilde{B}/2}\Big[ -\me^{2 C}\Big( \frac{\dd t}{r}+ \alpha \eta + A_{(6)}\Big)^2 +\me^{2C/3} \Big(\dd y + \me^{2 C/3}\Big( \frac{\dd t}{r} + \alpha \eta + A_{(6)}\Big)\Big)^2+\frac{\dd r^{2}}{r^2}\nonumber\\[4pt]
&+\eta^2+\me^{\widetilde{B}-2C/3} \,\dd s^{2}_{6}\Big]\, .
\end{align}
If we take $C=\alpha=A_{(6)}=0$, the first line gives precisely the metric for AdS$_3$ written as a U$(1)$ fibration over AdS$_2$. The effect of a non-trivial scalar $C$ and connection pieces $\alpha,A_{(6)}$ is to make the black string rotate. Note that this is precisely the form of the near-horizon of the black string found in \cite{Hosseini:2019lkt} uplifted to a 10d solution of Type IIB. The fluxes consist of an axio-dilaton and five-form flux given by
\begin{align}
C^{\text{IIB}}_{0}+\ii \me^{-\phi_{\text{IIB}}} &=\tau_{1}+ \ii \tau_{2}\, ,\\
F^{\text{IIB}}_5&= (1+\star_{10})\,\dd \bigg[ c_3 \wedge \Big(\dd y + \me^{2C/3} \Big(\frac{\dd t}{r}+\alpha \eta + A_{(6)}\Big)\Big) \nonumber\\[4pt]
&\phantom{=}\;\,- \me^{-\widetilde{B}+4C/3} \Big(\frac{\dd t}{r}+\alpha \eta + A_{(6)}\Big) \wedge \Big( \eta \wedge \frac{\dd r}{r}+\me^{\widetilde{B}-2C/3} J_{(6)}\Big) \wedge \dd y\bigg]\, .
\end{align}

Having given the metric and fluxes we now specify the supersymmetry conditions that the geometry must satisfy. These can be derived from the 11d supergravity ones by reducing them on the torus. Note that the cone appearing in the metric in \eqref{IIBcone} has an SU$(4)$ structure which is inherited from the SU$(5)$ structure of our 11d solutions. We denote the corresponding two-form by $j_{(8)}$, and we can decompose it as
\be
j_{(8)}= r \eta \wedge \dd r + r^2 \me^{\widetilde{B}-2C/3}\, J_{(6)}\, ,
\ee
where $J_{(6)}$ is the two-form that we found in the decomposition \eqref{decompJ6}. This two-form corresponds to the balanced SU$(3)$ structure of the 6d space. On this SU$(3)$ structure, we previously found the conditions:
\begin{align}
\dd J_{(6)}^2 &= 0 \,,\\
\dd\Omega_{(6)} &= \ii\big( P + \tfrac{1}{3}\dd^c C - Q \big) \wedge \Omega_{(6)} \,.
\end{align}
The geometry must in addition satisfy the Bianchi identities and Maxwell equations that we discussed earlier in this section subject to the potential $c_3$ satisfying 
\be
j_{(8)}\lrcorner \, \dd c_3=2 r^{-3}\me^{-\widetilde{B} + 4C/3}\dd a^{(1,1)} \, .
\ee
The first of the Maxwell equations \eqref{MaxwellonT2} is the master equation, which can be rewritten as
\begin{equation}
\begin{aligned}
&\me^{2C/3}\,\Box_6\big(\me^{-2C/3} \big(R_{C(6)} + \tfrac{2}{3} \square_6 C-\tfrac{1}{2\tau^2_{2}} \,|\dd \tau|^2 \big)\big) - \tfrac{1}{2}\big(R_{C(6)} + \tfrac{2}{3} \square_6 C-\tfrac{1}{2\tau^2_{2}} \,|\dd \tau|^2 \big)^2 \\
&+2 \big| \rho_{(6)} - \tfrac{1}{3}\dd\dd^cC + \dd Q \big|^2 \,=\, 0 \,.
\end{aligned}
\end{equation}
Note that the master equation is independent of the fluxes here. Further, notice that the conditions reduce to those of \cite{Kim:2005ez} if one sets $C=\alpha = A_{(6)} =c_3=0$.

The solutions in this classification may be interpreted as the near-horizon geometries of rotating black strings. When one inserts a Riemann surface into the balanced 6d base it is natural to interpret these as arising from the compactification of rotating D3-branes on the Riemann surface. Moreover, this is not the most general setup that can be considered and it would be interesting to further investigate extensions. A possible method for doing this is to reduce the 11d setup studied here on a torus which is also fibered over the AdS$_2$, as we alluded to at the beginning of this section. This will necessarily lead to two free constants in the Type IIB solution and also to more general fluxes. However, such solutions are far more involved than the ones presented in this section.


\section{Conclusions and future directions}\label{sec:conclusion}

In this paper we studied the geometry of supersymmetric solutions which may be interpreted as the near-horizon of rotating black holes and strings embedded in 11d supergravity and Type IIB respectively. This generalizes the results of \cite{Kim:2006qu} and \cite{Kim:2005ez}. Due to the generality of our ansatz the black holes covered by our classification can include both electric and magnetic flavour fluxes and angular momentum when viewed from 4d.  Note that this does not translate into magnetic fluxes in 11d but rather into fibrations of the manifold.\footnote{The role of baryonic symmetries is slightly more mysterious but we believe that these should also be covered by our work.} Similar statements apply for the 5d black strings in Type IIB that we considered.

One natural extension of our work is to consider a more general classification of the black strings in Type IIB. In performing the duality chain we aimed for a simplified solution consisting of only five-form flux and axio-dilaton. One could in fact include a complex three-form flux in the setup. This may be achieved from 11d by allowing the flux components of $\tilde{h}^{(2,2)}$ to have legs along the torus directions. The minimal extension would be adding in a transgression term of the form discussed in \cite{Donos:2008ug}, however we expect that one can be more general by also allowing for rotation along the $T^2$-directions. A preliminary analysis showed that this case is rather involved with all fluxes turned on and a non-holomorphic axio-dilaton. For the sake of presentation we have given only the simpler case. 
 
It would be interesting to formulate an extremization principle for these geometries along the lines of \cite{Couzens:2018wnk}. This seems quite challenging though there are glimpses of hope. The entropy of the black hole and string can be seen to be given by the same formula as in the non-rotating case. In particular the actions presented in section \ref{ssec:susyaction} reduce to simple integrals \eqref{SBHSUSY} and \eqref{SBSIIB} which can easily be computed in the toric case. The difficulty arises in evaluating the integrals which impose flux quantisation. One should be able to compare with the field theory results in \cite{Hosseini:2019iad,Hosseini:2019lkt} for rotating black holes and black strings. We have preliminary results on this extremization problem and plan to present these in the future. 

Some alternative and intriguing avenues are to attempt to perform a similar analysis for Euclidean black saddles \cite{Bobev:2020pjk}, for other rotating black hole solutions and to include higher derivative corrections \cite{Bobev:2020egg,Bobev:2020zov,PandoZayas:2020iqr}. There are many results with which one could compare for black holes in other theories, for example \cite{Lanir:2019abx,Cabo-Bizet:2018ehj,Cabo-Bizet:2020nkr,Benini:2020gjh,Lezcano:2019pae,Cassani:2019mms,Hosseini:2017fjo,Hosseini:2017mds}. It would also be desirable to understand the connection with Sen's entropy function \cite{Sen:2005wa,Morales:2006gm,Ghosh:2020rwf} and whether one can perform a similar classification for near extremal black holes \cite{David:2020jhp,Larsen:2020lhg,Larsen:2019oll}.

\subsection*{Acknowledgments}
It is a pleasure to thank Stefan Vandoren and Thomas Grimm for useful discussions.  C.C. would like to thank Jerome Gauntlett, Dario Martelli and James Sparks for a previous collaboration on a related topic. C.C. acknowledges the support of the Netherlands Organization for Scientifc Research (NWO) under the VICI grant 680-47-602 and E.M. acknowledges support by the FOM programme “Scanning New Horizons” .  

\appendix


\section{Black hole near-horizons and observables}\label{app:NH}

In this appendix we will study the general form of the near-horizon of a black hole. This analysis serves two purposes. Firstly it will motivate the ansatz we take in section \ref{sec:ansatz} for the 11d supergravity solution, in particular the warping of the metric and the temporal fibration. Despite this, in the main text we will use a more general ansatz to the one motivated here purely for convenience of the notation. It is understood that one must impose an additional constraint on the geometry in order for it to be the near-horizon of a black hole as we will show later in this section. 

The second purpose for this analysis is to determine how to evaluate the physical observables for our solution. The parametrization of the metric which is most useful for obtaining the conditions arising from supersymmetry is not the one that is most useful for defining the observables such as the entropy and angular momentum of the black hole where an explicit AdS$_2$ factor is used. The analysis of this section will allow us to translate between the two view-points and compute observables easily from the form of the metric obtained from supersymmetry.

\subsection{General near-horizon metric}

The general form of the metric for the near-horizon of a rotating black hole is \cite{Lucietti:2012sa}\footnote{We have made some trivial redefinitions to the form of the metric appearing in \cite{Lucietti:2012sa}, in particular we have changed coordinates on AdS$_2$ from Gaussian Null coordinates to Poincar\'e coordinates and extracted an overall factor from each of the sub metrics.} (see also \cite{Kunduri:2013gce} and references therein)
\be\label{AdS2NH}
\dd s^2 = \Gamma(y)\Big[ -r^2 \dd t^2 +\frac{\dd r^2}{r^2} +G_{MN}(y) \dd y^M \dd y^N+\gamma_{\mu\nu}(y)( \dd \phi^{\mu}+k^{\mu} r \dd t)( \dd \phi^{\nu}+k^{\nu} r \dd t)\Big]\, .
\ee
Here $\phi$ are periodic coordinates and $k^{\mu}$ are \emph{constants} related to the near-horizon value of the chemical potentials of the angular momentum of the black hole. The functions of the metric all depend on the $y$ coordinates and are independent of the $\phi$'s. We do not need to specify the ranges of the indices $M$ and $\mu$ for the argument but let the range of $\mu\in \{1,\ldots, n\}$\footnote{Note that $n$ cannot be zero otherwise the black hole is not rotating and we fall into the class of solutions given in \cite{Kim:2006qu}.}. Note that the first two entries of the metric are precisely the metric on AdS$_2$ with unit radius. Moreover it is clear from this form that there is an SO$(2,1)\times \text{U}(1)^n$ isometry.\footnote{The SO$(2,1)$ algebra of the metric in these coordinates is realised by the three Killing vectors
\begin{equation*}
H= \partial_{t}\, ,\quad D = t \partial_{t}+ r \partial_{r}\, ,\quad K= (t^2+r^{-2})\partial_{t}-2 t r \partial_{r} - 2 r^{-1} k^{i} \partial_{\psi^{i}}\, ,
\end{equation*}
where the $\psi_{i}$ denote the U$(1)$ symmetries of the internal manifold. Note that the generators are twisted with respect to the U$(1)$ symmetries of the internal manifold which are gauged over the AdS$_2$. It is important that the twisting parameters, the $k^{i}$'s are constant otherwise the SO$(2,1)$ algebra is broken. The Killing vectors satisfy the algebra
\begin{equation*}
[H,D]=H\, ,\quad [K,D]=-K\, ,\quad [H,K]=2 D\, 
\end{equation*}
which is precisely the algebra of the conformal group in 1d and commutes with the isometries of the internal manifold.}

The metric in this form is useful for computing the observables of the black hole however it is not as useful when trying to impose supersymmetry. Due to the gauging over AdS$_2$ it is finicky to try to implement SUSY preservation in this form. It is known that supersymmetry in 11d supergravity imposes that a metric admits either a timelike or null Killing vector \cite{Gauntlett:2002fz, Gauntlett:2003wb}. Since the form of the metric we are considering above has a time-like Killing vector we will focus on this case\footnote{One could also have attacked the problem using the null Killing vector of AdS$_2$. The benefit of using the timelike Killing vector is that it is transferable to the case of black strings in Type IIB and so we pursue this choice here.}. It is then useful to rewrite the metric so that the timelike Killing vector is manifest. This will lead to the time-direction being fibered over the remaining directions. A small rearrangement puts the metric into the form
\begin{align}
\dd s^2=&\Gamma(y) \bigg[ -(1- \gamma_{\tau\kappa}k^{\tau}k^{\kappa}) \Big(r \dd t -\frac{k^{\mu}\gamma_{\mu\nu}\dd \phi^{\nu}}{1-\gamma_{\sigma\rho}k^{\sigma}k^{\rho}} \Big)^2 + \frac{\dd r^2}{r^2}+ G_{mn}(y) \dd y^{m} \dd y^{n}\nonumber\\
&+\Big( \gamma_{\mu\nu} + \frac{k^{\sigma}\gamma_{\sigma \mu} k^{\rho}\gamma_{\rho \nu}}{1-\gamma_{\kappa \tau}k^{\tau}k^{\kappa}}\Big) \dd \phi^{\mu}\dd \phi^{\nu}\bigg]\, .\label{AdS2NHclass}
\end{align}
The metric now exhibits the timelike Killing vector in a simple form. It is then natural to take as ansatz\footnote{We change the radial coordinate as $r \rightarrow r^{-1}$ in order to write the transverse directions to the timelike foliation as a cone in the main text.}
\be\label{app:Ansatzeq}
\dd s^2 = \me^{2 B} \Big[ - \me^{2C} \Big( \frac{\dd t}{r} +\hat{A}\Big)^2+ \frac{\dd r^2}{r^2} + \dd s^{2}_{9}\Big]
\ee
for the near-horizon, with $\hat{A}$ an $r$-independent one-form on the 9d base. In this rotated form the AdS$_2$ factor is obscured, however as we mentioned previously this form is far more amenable to imposing supersymmetry.
However this ansatz does come with some downsides. Firstly computing observables, such as the horizon area are not nearly as clear as in the form given in the ansatz \eqref{AdS2NH}. Moreover it is not clear which solutions can be identified with the near-horizon of a rotating black hole from the form in \eqref{app:Ansatzeq}, in particular the scalar $C$ is arbitrary in our ansatz whilst its analogue in \eqref{AdS2NHclass} is constrained. We shall study this constraint shortly however in the main text we shall refrain from imposing it for as long as possible. We will see that we can proceed unabated in the classification without needing to impose such a condition.

\subsection{Constraints from the near-horizon}

In this section we shall look at the additional constraints imposed on the metric ansatz used in the main text which follow from it being the near-horizon of a black hole. We shall compare our ansatz with the general form of the near-horizon given in the previous section, rewriting the expressions in terms of quantities adapted to the metric in the form of the classification. The classification implies that the metric takes the form
\be
\dd s^2=\me^{2B}\Big[ -\me^{2C}\Big(\frac{\dd t}{r}+\alpha \eta + A\Big)^{2}+ \frac{\dd r^2}{r^2} + G_{mn}(y) \dd y^{m} \dd y^{n}+ g_{\mu\nu} \dd \phi^{\mu} \dd \phi^{\nu} \Big]
\ee
where we have written the metric with the same splitting as earlier. It is trivial to identify
\begin{align}
\me^{2B}= \Gamma(y)\, ,\quad \me^{2C}= 1- |k|^{2}_{\gamma}\, ,\quad -\me^{-2 C} k^{\mu} \gamma_{\mu\nu} \dd \phi^{\nu}= \alpha \eta + A\, ,\quad  g_{\mu\nu}= \gamma_{\mu\nu} + \frac{k^{\sigma}\gamma_{\sigma \mu} k^{\rho}\gamma_{\rho \nu}}{1-\gamma_{\kappa \tau}k^{\tau}k^{\kappa}}\, .
\end{align}
Note that we have defined $|\cdot|_{\gamma}$ to be the norm with respect to the metric $\gamma$, similarly we let $|\cdot|_{g}$ denote the norm with respect to $g$. Simple manipulations of these definitions gives
\begin{align}
g_{\mu\nu} k^{\nu} = \me^{-2 C} \gamma_{\mu\nu} k^{\nu}\, ,\quad |k|_{g}^2=\me^{-2 C} |k|_{\gamma}^2\, , \quad \me^{2 C}= (1+ |k|^{2}_{g})^{-1}\, .
\end{align}
Note that this implies we can constrain the scalar $C$ in terms of data of the fibration, in particular
\be
\me^{-2 C}= 1+|\alpha \eta + A|^{2}_{9}\, .
\ee
Finally rewriting this in terms of the full metric of the classification we find the condition
\be
\me^{-2 C}= 1+ \alpha^{2}+ \me^{3B+ C/3} |A|^2
\ee
where the final norm is with respect to the metric on the balanced manifold. Let us further analyse the condition on the fibration in the time-direction. We have
\be
\alpha \eta + A=- k^{\mu} g_{\mu\nu} \dd \phi^{\nu}\, .
\ee
Therefore in order to specify $\alpha$ and $A$ we should specify $\eta$, the metric $g_{\mu\nu}$ and a set of constants $k^{\mu}$. These constants $k^{\mu}$ are related to the near-horizon values of the chemical potentials of the angular momentum of the black hole (when viewed from 11d). As a final step let us rewrite the metric used in the arguments above so that the R-symmetry vector is manifest. We want to identify
\be
G_{mn} \dd y^{m} \dd y^{n}+ g_{\mu\nu} \dd \phi^{\mu} \dd \phi^{\nu}\equiv (\dd z+P)^2 +\me^{D} \dd s^{2}_{8}\, .
\ee
Clearly the $G_{mn}$ part fits in trivially after extracting out the required warp factor. The angular part can be written as
\begin{align}
g_{\mu\nu} \dd \phi^{\mu} \dd \phi^{\nu}&= g_{zz} \dd z^2+ 2 g_{z \hat{\mu}}\dd z \dd \phi^{\hat{\mu}}+ g_{\hat{\mu}\hat{\nu}} \dd z^{\hat{\mu}}\dd z^{\hat{\nu}}\\
&= (\dd z+ g_{z \hat{\mu}}\dd \phi^{\hat{\mu}})^{2}+ (g_{\hat{\mu} \hat{\nu}}-  g_{z \hat{\mu}} g_{z \hat{\nu}})\dd \phi^{\hat{\mu}}\dd \phi^{\hat{\nu}}
\end{align}
where we have used that $g_{zz}=1$ and we should identify $g_{z \hat{\mu}}\dd \phi^{\hat{\mu}}= P$. Therefore we have
\be
\dd s^{2}= (\dd z+P)^2+ \me^{D} \dd s^{2}_8= (\dd z+P)^2 + G_{mn} \dd y^{m} \dd y^{n} +(g_{\hat{\mu} \hat{\nu}}-P_{\hat{\mu}}P_{\hat{\nu}})\dd \phi^{\hat{\mu}} \dd \phi^{\hat{\nu}}\, .
\ee
Inserting the decomposition into the connection piece of the timelike fibration we have
\be
k^{\mu}g_{\mu\nu} \dd \phi^{\nu}=k^{z} \dd z + k^{\hat{\mu}} g_{\hat{\mu}z} \dd z+ k^{\hat{\mu}}g_{\hat{\mu}\hat{\nu}} \dd \phi^{\hat{\nu}}+ k^{z} g_{z \hat{\mu}}\dd \phi^{\mu}
\ee
from which we find
\be\label{alpha:appexplicit}
-\alpha (\dd z+P) -A=k^{\mu}g_{\mu\nu} \dd \phi^{\nu}=(k^{z}+k^{\hat{\mu}}P_{\hat{\mu}} )( \dd z+ P)+  ( g_{\hat{\mu}\hat{\nu}}-P_{\hat{\mu}}P_{\hat{\nu}}) k^{\hat{\mu}}\dd \phi^{\hat{\nu}}\, .
\ee
Therefore given a vector of constants parametrising the rotation and the internal metric one can construct $\alpha \eta+ A$. In fact if one imposes that the internal manifold is toric one may write the gauge field in a simple way as we have explained in section \ref{Toricpart}.

\subsection{Observables}

Let us now use the near-horizon solution to study what observables we can compute. The three main observables are the entropy of the black hole, the angular momentum and its electric/magnetic charges, all of which can be computed in the near-horizon. One may also ask if it is possible to compute the electrostatic potential and angular velocity, however these observables require some knowledge of the UV data since they are defined as
\be 
\mathcal{O}_{\text{BH}}= \mathcal{O}_{\text{NH}}- \mathcal{O}_{\infty}\, .
\ee
In this section we will focus on rephrasing the computation of the entropy, electric charges and angular momentum in terms of integrals over various cycles of the internal manifold.

\subsection*{Entropy}

First consider the entropy of the black hole. The entropy is given up to normalization by the area of the horizon of the black hole. In order to compute the horizon area one should write the metric so that a bona-fide AdS$_2$ factor appears in the metric and the internal manifold is fibered over this. Clearly in order to compute the entropy in this way the metric of use to us is the one given in \eqref{AdS2NH} and not the one that naturally comes out from supersymmetry. 
With this rewriting the horizon manifest and the entropy is given by 
\be
S_{\text{BH}}=\frac{1}{4 G_{2}}\, ,
\ee
where the Newton's constant is that of a 2d theory admitting the AdS$_2$ near-horizon as a vacuum solution. In order to compute the Newton's constant (at leading order, we will not make any comments about subleading corrections though these are certainly very interesting) we should look at reducing the 11d Einstein-Hilbert term of 11d supergravity on the AdS$_2$ background in \eqref{AdS2NH}. We have\footnote{To save cluttering the notation we let $\dd y \wedge \dd \phi$ denote $\bigwedge \dd y^{m} \wedge \bigwedge \dd \phi^{\mu}$.}
\be
\frac{1}{G_{11}}\int_{M_{11}} R \dd \vol_{M_{11}}= \frac{1}{G_{11}}\int_{\M_{2}}R_{2}\dd \vol_{2} \int_{Y_9}\Gamma(y)^{\tfrac{9}{2}} \sqrt{\det(G)}\sqrt{\det(\gamma)}\dd y \wedge \dd \phi \equiv \frac{1}{G_{2}}\int_{\M_{2}} R_{2}\dd \vol_{2}\, ,
\ee
from which we identify
\begin{equation}
\begin{aligned}
\frac{1}{G_{2}}&= \frac{1}{G_{11}} \int_{Y_9}\Gamma(y)^{\tfrac{9}{2}} \sqrt{\det(G)}\sqrt{\det(\gamma)}\dd y \wedge \dd \phi\, \\
&= \frac{1}{G_{11}} \int_{Y_{9}} \Gamma(y)^{\tfrac{9}{2}} \dd \vol_{9}\, .
\end{aligned}
\end{equation}
Let us now translate this result into the notation of the metric arising from supersymmetry, namely \eqref{AdS2NHclass}. We expect that the difference is precisely a warping of the volume form which indeed turns out to be the case. To this end let us compute the volume of the internal manifold. We distinguish between the two volume forms by writing $\dd \vol_{\text{SUSY}}$ for the volume form in the form natural from supersymmetry. We have
\be
\dd \vol_{\text{SUSY}}= \Gamma(y)^{\tfrac{9}{2}} \sqrt{\det (G)} \sqrt{\det\Big( \gamma_{\mu\nu} + \frac{k^{\sigma}\gamma_{\sigma \mu} k^{\rho}\gamma_{\rho \nu}}{1-\gamma_{\kappa \tau}k^{\tau}k^{\kappa}}\Big)} \dd y \wedge \dd \phi \, .
\ee
We can expand the determinant second determinant. Using the fact that for an invertible matrix $A$ and vectors $v,w$ one has
\be
\det(A+ v w^T)= \det(A)( 1+ w^T A^{-1} v)\, ,
\ee
we have
\begin{align}
\det\Big( \gamma_{\mu\nu} + \frac{k^{\sigma}\gamma_{\sigma \mu} k^{\rho}\gamma_{\rho \nu}}{1-\gamma_{\kappa \tau}k^{\tau}k^{\kappa}}\Big)&= \det(\gamma_{\mu\nu})\Big( 1+ \frac{1}{1-\gamma_{\kappa \tau}k^{\tau}k^{\kappa}} k^{\sigma}\gamma_{\sigma \mu}\gamma^{\mu\nu} k^{\rho}\gamma_{\rho \nu}\Big)\nonumber\\
&= \frac{\det(\gamma_{\mu\nu})}{1-\gamma_{\kappa \tau}k^{\tau}k^{\kappa}}\, .
\end{align}
It follows that 
\begin{align}
\dd \vol_{\text{SUSY}}&=\frac{\Gamma(y)^{\tfrac{9}{2}} \sqrt{\det(G)}\sqrt{\det(\gamma)}}{\sqrt{1-\gamma^{\kappa}\gamma^{\tau}\gamma_{\kappa \tau}}}\dd y \wedge \dd \phi\, ,
\end{align}
and therefore
\be
\frac{1}{G_2}= \int_{Y_9} \Gamma(y)^{\tfrac{9}{2}}\sqrt{1- \gamma_{\kappa\tau}k^{\kappa}k^{\tau}}\dd \vol_{\text{SUSY}}\, .
\ee
Our proposal for computing the entropy is therefore
\begin{equation}
\begin{aligned}
S_{\text{BH}}=&\frac{1}{4 G_{11}}\int_{Y_{9}} \me^{-3 B - C/3} \eta \wedge \frac{J^4}{4!}\\
=&\frac{1}{4 G_{11}}\int_{Y_{9}}\eta \wedge \dd \eta \wedge \frac{J^3}{3!}\label{BHentropy}\, ,
\end{aligned}
\end{equation}
where we used \eqref{Jintodeta} in the final equality. As discussed in section \ref{ssec:susyaction} this is precisely the same formula as the entropy in the non-rotating case. One should view this section as a proof that the quantity computed in section \ref{ssec:susyaction} really is the entropy of the black hole.


\subsection*{Electric charges}

Next let us consider the quantization of the four-form flux which will give rise to the electric charges of the theory. In the presence of a Chern--Simons term there is more than one definition of a charge. One can consider the gauge-invariant but non-conserved charge
\be
Q=\frac{1}{(2\pi \lp)^6}\int_{\Sigma_{7}} \ast_{11} G_4\, ,
\ee
where we integrate over all compact seven-cycles of the geometry. Alternatively the Page charge
\be
Q=\frac{1}{(2\pi \lp)^6} \int_{\Sigma_{7}} \Big(\ast_{11} G_4+ \frac{1}{2} C_3 \wedge G_4\Big)\, ,
\ee
is conserved by application of the Maxwell equation but is not gauge invariant due to the bare potential appearing in the definition. In the following we will consider only the Page charge since it defines a conserved charge. In order to be able to write this charge we must be able to at least locally write the four-form flux in terms of a potential three-form. This is equivalent to the requirement that $\tilde{h}^{(2,2)}$ as defined in \eqref{h22def} can be written (at least locally) in terms of a potential. In fact, if we demand that it is exact, i.e. that the potential is a globally defined three-form, it follows that there is no M5-brane charge. Substituting our ansatz into the Page charge we find
\begin{align}
Q= \frac{1}{(2 \pi \lp)^6} \int_{\Sigma_{7}}\eta \wedge\bigg[& \me^{-2C/3}  \dd\eta \wedge \frac{J^2}{2}-\frac{1}{2} \bigg( C^{(2)}\wedge H^{(2,2)} + C^{(3)} \wedge \dd C^{(2)} \nonumber\\
&+ \me^{2C/3} J \wedge \Big(C^{(3)}\wedge \dd \alpha - C^{(2)} \wedge (\alpha \dd \eta + \dd A)\Big)\bigg)\bigg]\, ,
\end{align}
where we have introduced the potentials
\begin{align}
H^{(1,1)}&=\dd C^{(1)}\, ,\nonumber\\
\ii(H^{(2,1)}-H^{(1,2)})&= \dd C^{(2)}\, ,\\
H^{(2,2)}&=\dd C^{(3)}= \dd \eta \wedge C^{(2)}\, .\nonumber
\end{align}

\subsection*{Angular momentum}

We now want to find a similar formulation for computing the angular momentum of the black hole. To such an end we may use the results of \cite{Katmadas:2015ima}, (see also \cite{Hanaki:2007mb} for the analogous computation for 5d black rings), which gives the formula for computing the Komar integral for the Noether current of a Killing vector, $\xi$ in 11d supergravity. By an abuse of notation we will also call the dual one-form $\xi$. The angular momentum is then given by
\begin{align}
J_{\xi}=\frac{1}{\S_{\text{SUSY}}} \int_{Y_9} \bigg[ \ast_{11} \dd \xi + ( \xi\cdot C_3 ) \wedge \ast_{11} G_4 + \frac{1}{3} ( \xi \cdot C_3) \wedge C_3 \wedge G_4\bigg]
\end{align}
where the three-form potential $C_3$ should be chosen so that it has vanishing Lie derivative along the given isometry. 
Since this formula is dependent on the choice of Killing vector we will refrain from writing this more explicitly and just include it for completeness.


\bibliographystyle{JHEP}

\bibliography{ADSCFT}

\end{document}